\documentclass[aps,preprint,amssymb]{revtex4}

\usepackage{epsfig}

\begin{document}
\renewcommand{\thefootnote}{\fnsymbol{footnote}} 
\newcommand{\be}{\begin{equation}}
\newcommand{\ee}{\end{equation}}
\newcommand{\vv}[1]{\mbox{\boldmath $#1$}}
\newcommand{\vtimes}{\vv{\times}}
\newcommand{\E}{\varepsilon}                   
\newcommand{\EL}{E^{\rm Lan}}                  
\newcommand{\eL}{\varepsilon^{\rm Lan}}        
\newcommand{\Heff}{H_{\rm eff}}                
\newcommand{\me}{m_{\rm e}}                    
\newcommand{\mat}{m_{\rm at}}                  
\newcommand{\He}{H_{\rm e}}                    
\newcommand{\Veff}{V_{\rm eff}}                
\newcommand{\rL}{r_{\rm L}}                    
\newcommand{\rc}{r_{\rm c}}                    
\newcommand{\RL}{R_{\rm L}}                    
\newcommand{\Rc}{R_{\rm c}}                    
\newcommand{\cL}{\chi_{\rm L}}                 
\newcommand{\cc}{\chi_{\rm c}}                 
\newcommand{\Bc}{B_{\rm crit}}                 
\newcommand{\XL}{X_{\rm L}}                    
\newcommand{\YL}{Y_{\rm L}}                    
\newcommand{\Xc}{X_{\rm c}}                    
\newcommand{\Yc}{Y_{\rm c}}                    
\newcommand{\xL}{x_{\rm L}}                    
\newcommand{\yL}{y_{\rm L}}                    
\newcommand{\xc}{x_{\rm c}}                    
\newcommand{\yc}{y_{\rm c}}                    
\newcommand{\qc}{q_{\rm c}}                    
\newcommand{\pc}{p_{\rm c}}                    
\newcommand{\Lt}{{\cal L}}                     

\title{Magnetically induced anions}

\author{Victor G. Bezchastnov} 
\email[]{E-mail: Victor.Bezchastnov@tc.pci.uni-heidelberg.de}
\thanks{Permanent address: {\em Department of Theoretical Astrophysics, 
        Ioffe Physical-Technical Institute, 194021 St.-Petersburg, Russia}} 

\author{Peter Schmelcher} 
\email[]{E-mail: Peter.Schmelcher@tc.pci.uni-heidelberg.de}

\author{Lorenz S. Cederbaum}
\email[]{E-mail: Lorenz.Cederbaum@urz.uni-heidelberg.de}

\affiliation{Theoretische Chemie, Physikalisch-Chemisches Institut,
             Universit\"at Heidelberg, INF 229, D-69120 Heidelberg, Germany}

\date{\today}

\begin{abstract}
\noindent 
The main focus of this review is on magnetically induced anions. Before discussing these new 
anionic states which exclusively exist in the presence of magnetic fields, we review in some 
detail the anionic physics without external field. For completeness, we also outline the properties 
of field-free existing anions when exposed to magnetic fields. The magnetically induced states
constitute an infinite manifold assuming that the nucleus of the anion is infinitely heavy. 
At laboratory field strengths the corresponding binding energies show two different scaling 
properties belonging to the ground ($\propto B^2$) and the excited ($\propto B^3$) 
magnetically induced states. We provide a detailed discussion of the physics of the moving anion 
taking into account the coupling between the anionic centre of mass and its electronic degrees 
of freedom. A number of field-adapted techniques exploiting exact constants of motion and the 
adiabatic separation of motions are applied to simplify the Hamiltonian that describes the 
effective interaction between the centre of mass and the electronic degrees of freedom. 
Employing classical simulations the autodetachment process of the anions in the field is observed 
and a rich variety of spectral properties of moving anions is predicted: depending on the parameters, 
such as the mass and polarizability of the neutral system as well as the field strength, 
induced bound states, resonances and detaching states of the corresponding anions are to be expected. 
An outline of an {\it{ab initio}} quantum approach is provided that allows quantum dynamical 
investigations of the moving anion. 
\end{abstract}

\maketitle

\section{Introduction}

The behavior and properties of negative ions have been the subject of intense 
research for many decades. There exists a vivid interest in the electronic structure and
dynamics of negative ions, both from the theoretical and experimental points of view 
(see, e.g., \cite{Massey, Scheller_review}, and the references therein). 
Negative ions play an important role in stellar and terrestrial atmospheres 
as well as in laboratory and cosmic plasmas \cite{Gray1992, Chutjian1996}. 
The main processes responsible for the radiative 
properties of the solar photosphere are the bound-free transitions of the anion H$^-$, 
which have been investigated theoretically in Refs.~\cite{Wildt1939, Chandrasekhar1958}. 
The same ion could play a crucial role in the production of the interstellar H$_2$ molecules. 
Beyond this negative ions are of major relevance to chemical reaction dynamics in general and 
frequently occur either as intermediates or final products of chemical reactions.

The formation and binding mechanisms of negative ions are of delicate nature. Therefore, 
sophisticated theoretical as well as spectroscopic tools have to be invented in order to prove 
their existence and probe their properties. Although research on negative ions is both theoretically 
and experimentally a well-established area, the past ten years have seen enormous progress and added 
a lot to our knowledge on anions, see, e.g., Ref.~\cite{Andersen_review} for a recent review and the 
references therein for other reviews. As an example of a new experimental method we mention the 
threshold photodetachment to excited states of the neutral atom in conjunction with resonance 
ionization detection that allows the precise determination of binding energies in weakly bound 
atomic negative ions (electron affinity less than $0.2$eV) \cite{Petrunin1995, Andersen1997}. 
From a theoretical point of view the {\it ab initio} description of anions employing e.g. 
multiconfiguration Hartree-Fock with relativistic corrections, multiconfiguration Dirac-Fock, 
relativistic coupled cluster or multireference configuration interaction has advanced substantially. 
In Section~II we will provide an outline of the binding mechanisms of various types of anions including 
atoms, molecules and to some extent also clusters thereby discussing the subtleties of their existence 
and spectral properties. Most neutral atomic and small molecular species lead to anions with a few 
bound states. In some cases even no anionic counterpart exists.

External fields add significantly to the variety of anions as well as to the richness of their properties. 
Focusing on magnetic fields a key work on anions exposed to homogeneous fields was done by 
Avron, Herbst and Simon in 1981 \cite{Avron}. Applying tools from functional analysis they arrived 
at the surprising statement that any anion possesses an infinite number of bound states for arbitrary 
magnetic field strength. Motivated by this spectacular discovery the authors of this review performed 
recently a number of investigations on the physical properties of these newly formed anions. 
In particular, they went beyond the assumption of an infinitely heavy nucleus that was applied in 
Ref.~\cite{Avron}. The main concern of the present review is to summarize and elucidate the 
state-of-the-art concerning anions exposed to external magnetic fields with particular emphasis on the 
so-called magnetically induced anions. The latter states and/or species exist exclusively due to the 
presence of the external magnetic field. In Section~III we discuss anions in the presence of a magnetic 
field dealing with the response of well-known anionic states to the field. Section~IV is dedicated to 
the discussion of magnetically induced anions. First we will review the case of a static anion assuming 
infinite anionic mass. Thereafter we proceed to include effects due to the collective motion of the 
anions that possess severe impact on their spectral and dynamical properties. In particular we draw 
attention to classical simulations which show that, depending on the relevant parameters, the 
motion-induced dynamics can lead to a decay of the magnetically induced states. An outline of a 
possible quantum approach to the moving magnetically induced anions is given. In section V we
provide some brief conclusions.

\section{Anions in field-free space.}

The major part of the theoretical and experimental studies on anions concerns their properties in 
field-free space. A variety of anionic states, e.g., ground, excited and resonance 
states of different atomic, molecular and cluster species is known. In order to 
classify these states they can be sorted with respect to the underlying mechanisms that support 
the attachment of an extra electron to a neutral species. Basically, one can divide 
these mechanisms into 
\begin{list}{}{}
\item A) valence binding, 
\item B) non-valence binding,
\item C) combined or mixed binding
\end{list}
mechanisms. In this classification the anionic {\it valence-bound states} are formed by accommodating 
an extra electron in an outer electronic shell of the neutral counterpart. 
The electron then becomes bound exclusively due to short-range attractive valence forces. 
In contrast, for the {\it non-valence-bound states} the extra electron is loosely bound in an orbital 
whose spatial extension exceeds significantly the extensions of the electronic orbitals of the neutral 
species. For such states, long-range attractive forces influence the motion of the extra electron and 
are the key ingredients of its binding. Finally, the {\it combined or mixed binding} happens if, e.g., 
a few outer electrons of the neutral species extend their spatial orbitals to accommodate 
the extra electron, or the external electron attaches to an excited Rydberg orbital of a molecule. 
For these states, the short-range valence forces support binding of the excess electron in 
the extended orbital, while the latter orbital is additionally stabilized by the long-range attraction 
to the inner part of the anion. Among the binding mechanisms introduced above only the valence binding 
is of pure character. The corresponding states can be referred to as {\it conventional anionic states}. 
For the non-valence binding, the short-range valence forces usually remain important as well. 
As a result, the long-range forces, although playing a dominant role, are in many cases not exclusively 
responsible for the stability and binding properties of the anion. A similar statement holds for 
the ``long-range part'' of the combined binding mechanisms. 
Rigorously speaking, the non-valence binding and the combined binding can be specified 
as the mechanisms where the short-range valence forces are not the 
{\it only or primary source} of supporting the anionic states, see, e.g., Ref.~\cite{Simons}. 
Another commonly used terminology is referring to the corresponding states as to 
{\it non-conventional anionic states}, see Ref.~\cite{Kalcher}.

In the following we present a brief overview of the anions in relation to the basic classes of the 
binding mechanisms specified above. We discuss some characteristic examples of 
the valence-bound, non-valence-bound and combined-bound anionic states. 
For each class, when appropriate, we address first atomic and then molecular or more complicated anions 
and consider examples of the ground, excited and resonance anionic states. 

\subsection{Valence-bound anions}

It is well-known that certain atoms are {\it unable} to form bound anions. Examples are 
He, Be, N, Ne, Mg, Ar, Mn, Zn, Kr, Cd, Xe, Hg, Rn. 
All other atoms can accommodate an extra electron in a {\it ground anionic state}, 
and the corresponding experimental and theoretical affinities are well documented, 
see, e.g., Refs.~\cite{Hotop1985, Andersen1999, Andersen_review}. 
For example, the hydrogen atom can attach an electron to form the 
H$^-$ ion within the electronic state $1s^2\,^{1}S_0$. 
The corresponding binding energy was theoretically calculated in Ref.~\cite{Pekkeris} 
as $0.754209(3)$~eV, and accurate measurements \cite{Lykke1991} 
yielded the experimental value $0.75420(2)$~eV. A search for excited bound electronic configurations 
has particularly been carried out in Ref.~\cite{Pekkeris} and led to a negative result. 
Later, a rigorous proof that the H$^-$ ion possesses only one
bound state was given in Ref.~\cite{Hill1977}. It is important to note that this state is stable 
exclusively due to correlation effects between the motions of the two electrons. 

It is only a few atoms, such as C and Si, that support {\it excited bound anionic states}. 
These states usually belong to the same symmetries, and therefore, 
exhibit in particular the same parities as the corresponding ground states.  
Also, a number of heavy atoms have been found to support bound stable excited anionic 
states. The corresponding theoretical calculations have to account for relativistic effects. 
Examples are the anions La$^-$, Lu$^-$, Bi$^-$ and Lr$^-$ \cite{Eliav}. 
Experimentally, excited states have been observed for La$^-$ \cite{Davis_in_press} 
and Lu$^-$ \cite{Covington1998}. The measured binding energies, $0.17$~eV and $0.16$~eV, 
respectively, turn out to be in good agreement with the theoretically predicted values. 

Many atomic anions possess {\it resonance states}. Often it is rather difficult 
to draw a definite conclusion if a state is an excited bound state or a resonance one. 
This is because the corresponding experimental observations can be of quite delicate nature, 
and the required theoretical calculations are usually rather involved, see, e.g. Ref.~\cite{Kalcher}. 
To provide an example let us consider the states $6s\,6p~^3P^{\rm o}_J$ of the Cs$^-$ ion that 
have been studied extensively. 
They form a manifold corresponding to the values $J=0,1,2$ of the total electronic momentum. 
On theoretical grounds, these states were identified in Ref.~\cite{Thumm1993} as resonances. 
The photodetachment studies presented in Ref.~\cite{Scheer1998} enabled an estimate of the 
position and width for the $^3P^{\rm o}_1$ state as $8$ and $5$~meV, respectively. 
The same studies however were not conclusive for the other states. 
Later, the positions of the resonances above the Cs threshold were computed 
\cite{Bahrim2000, Bahrim2001} as $9.11$, $21.84$ and $32.16$~meV with the respective 
widths of $5.78$, $19.33$ and $47.46$~meV. The fact that the energy bands corresponding to 
these resonance states strongly overlap elucidates the difficulty in resolving the individual 
states experimentally. 

The {\it valence-bound molecular anions} are formed by attaching an extra electron to a molecule 
with a {\it vacant} or {\it semi-filled} valence orbital. From the corresponding experiments, 
we know that bound anionic {\it ground states} are supported by almost all diatomic molecules. 
For example, in the anion CN$^-$ an extra electron occupies a non-bonding $\sigma$ orbital, while 
in the anion O$_2^-$ the added electron shares an anti-bonding $\pi^\ast$ orbital. Only a few diatomic 
molecules, in particular closed-shell ones such as H$_2$, N$_2$ and CO, do not support stable valence-bound 
anionic states. The latter two molecules, however can form combined-bound metastable anions, 
as described in section~C. Examples of more complex valence-bound anions can be found in Ref.~\cite{Simons}. 
They include, e.g., the anion H$_3$C-COO$^-$ with the extra electron occupying a delocalized $\pi$ orbital 
of the carboxylate group.

\subsection{Non-valence-bound anions}

The anions which one can call non-valence-bound are mainly molecular anions. 
From a theoretical point of view, following the methodology of the review~\cite{Simons}, 
some general remarks are appropriate. The latter concern the possible types 
of the long-range potentials experienced by an excess electron. 
Since the extra electron attaches to a neutral molecule, the latter cannot exert a 
Coulomb-like potential. A sort of ``exception'' are very large molecules, e.g., biomolecules 
comprising distinguishable cationic sites. Of course, they also cannot exert a 
Coulomb-like potential at large distances. However, if an excess electron approaches 
a certain spatial area around a cationic site it can become bound due to Coulomb 
attraction to this site. Such binding occurs in anions formed 
by attaching an extra electron to zwitterion tautomers H$_3$N$^{+}-$CHR$-$COO$^-$ 
with a well-defined cationic site H$_3$N$^+$ that are parts of fundamental biomolecules 
such as the amino acids. The binding energy of the zwitterion-derived 
anions is basically determined by a combination of attraction of the excess electron to 
the cationic site and repulsion exerted on this electron by the negatively charged end 
of the zwitterion. As pointed out in Ref.~\cite{Simons}, the longer the chain separating the 
two charged sites in the zwitterion, the stronger the net binding of an extra electron to 
the cationic site. 

For molecules that are not extraordinary large, the longest-range potential 
appears if the molecule possesses a permanent dipole moment $\vv{d}$. 
It is the {\it charge-dipole potential} 
\be
V_{\rm dip} = -\vv{d}\vv{r}/r^3,
\label{charge-dipole}
\ee
which scales with the distance $r$ between the electron and the neutral species as 
$V_{\rm dip} \propto r^{-2}$. A neutral species can also possess a permanent quadrupole 
moment, which is a tensor quantity $Q_{ik}$ ($i,k=1,2,3$ enumerate the Cartesian 
coordinate axes). Then an extra electron can experience the {\it charge-quadrupole potential} 
of the form 
\be
V_{\rm quad} = Q_{ik}x_ix_k/(2r^5), 
\label{charge-quadrupole}
\ee
where $x_i$ define the position vector, $\vv{r}=(x_1,x_2,x_3)$, of the attached 
electron with respect to the molecule, and the conventional tensor notations assume 
summations over the same indices. The latter potential scales with $r$ 
as $V_{\rm quad} \propto r^{-3}$. Other kinds of long-range potentials can appear due to 
the fact that the electric field of the external electron polarizes the neutral species 
thereby inducing dipole, quadrupole, and higher order momenta. This provides a 
contribution to the interaction energy of the electron and neutral species in form of 
{\it charge-induced-dipole}, {\it charge-induced-quadrupole} and higher-multipole induced 
potentials. The main term of the resulting {\it polarization potential} is 
\be
V_{\rm pol} = -\kappa_{ik}x_ix_k/(2r^6), 
\label{polarization}
\ee
where $\kappa_{ik}$ is the tensor of the (dipole) polarizability of the neutral species. 
The potential~(\ref{polarization}) scales with $r$ as $V_{\rm pol} \propto r^{-4}$. 

Since the charge-dipole potential is the longest-range potential which can be exerted by 
a neutral species, it is of major relevance in a variety of anions formed by attaching 
an electron to polar molecules. Some molecular anions are known to be formed due to 
the charge-quadrupole potential. The corresponding classes of anions are called the 
{\it dipole-bound anions} and {\it quadrupole-bound anions}, respectively. 
Below we briefly review some theoretical aspects related to these classes.

\paragraph{Dipole-bound anions.}

The charge-dipole potential~(\ref{charge-dipole}) has been rigorously examined regarding 
its ability to bind an electron. This potential is often called the {\it point dipole} potential. 
In Refs.~\cite{Fermi1947, Wightman1950} it was shown that if the magnitude 
of the dipole moment exceeds $1.625$~Debyes, the point dipole supports bound states of 
$\sigma$ symmetry. A dipole moment larger than $9.6375$~D is required to bind an electron in 
states of $\pi$ symmetry, while the bound states of $\delta$ 
symmetry can be supported if the dipole moment exceeds $24.218$~D. The binding energies 
are, however, infinite indicating that the electron ``falls into the attractive centre''. 
This feature, associated with a strong divergency of the potential~(\ref{charge-dipole}) 
at $r \to 0$, was a reason to consider the binding properties of the {\it fixed finite dipole}. 
The latter is a system of two static charges $q$ and $-q$ separated by a distance $R$ and providing 
the dipole moment of the magnitude $d=qR$. Being far away from the static finite dipole ($r \gg R$), 
an extra electron experiences the charge-dipole potential~(\ref{charge-dipole}). 
At smaller distances, the potential deviates from (\ref{charge-dipole}) and has a form of the Coulomb 
attractive/repulsion potential in the vicinity of the positive/negative static charge. 
In contrast to the point dipole, the binding energies of the finite dipole are finite 
for non-zero value of $R$. The first rigorous calculations of the energy levels, as functions of $R$, 
of an electron moving in the field of a finite dipole were presented in Ref.~\cite{Wallis1960}. 
Later, the binding properties the finite dipole were analyzed in 
Refs.~\cite{Crawford1967a, Crawford1967b, Brown1967}. The conditions for 
critical binding were shown to depend not on $q$ and $R$ separately, but on their product, 
i.e., on the value of the dipole moment $d$. Moreover, as discussed in Ref.~\cite{Turner1977}, 
the minimum magnitude of the dipole moment necessary to bind the electron in a $\sigma$ state 
is the same, $1.625$~D, as required for binding by the point dipole potential~(\ref{charge-dipole}). 
Both the potentials of a point dipole and of a fixed finite dipole were shown to support an infinite 
number of bound states of an electron. If the dipole is not fixed in space, i.e., is rotating, 
its binding properties weaken and the number of bound states is, at most, finite. 
In particular, a larger value of the dipole moment is needed to bind an electron. 
The critical dipole moment was found in Ref.~\cite{Garret1971} to increase with decreasing moment 
of inertia $I$ of the dipole and/or increasing total angular momentum $J$ of 
the system of the dipole and excess electron. 

Theoretical studies of electronic binding to polar molecules, which go beyond the simple point dipole 
and fixed finite dipole models, are well reviewed in Refs.~\cite{Desfrancois1995, Gutowski1998}. 
In particular, investigations of the binding of an electron to the diatomic molecule LiCl in 
Ref.~\cite{Jordan1976} revealed the finite dipole model predictions to be in poor agreement 
with the results of the {\it ab initio} calculations. 
In Ref.~\cite{Simons} a conclusion is drawn that there is a good reason to believe that in real situations 
the binding is due primarily to the dipole potential but, in no case can it be shown that the resultant 
anions are purely dipole-bound. The binding energy is 
determined not only by the dipole moment of the molecule but also by the nature of the molecule's 
other occupied orbitals as reflected in their Coulomb and exchange potentials. 
In some rigorous studies it was even possible to evaluate a percentage of binding resulting from 
other sources than the pure dipole attraction. For example, according to Ref.~\cite{Gutowski1996}, 
for the H$_3$C-CN$^-$ anion, which is assumed to be dipole-bound, in fact $53$~\% of the binding 
arises from the dispersion interaction between the excess electron and the other electrons. 
For those states where the charge-induced-dipole interaction plays an essential role in binding the 
electron, the average distance between this loosely bound electron and the polar molecule is typically 
$10-100$~\AA, see, e.g., Ref.~\cite{Smith1997}. 
Studying the binding of an electron to realistic polar molecules demonstrates the following fact: 
if a species forms a valence-bound anion, this does not preclude it from also forming a dipole-bound state. 
Advancing with respect to the theoretical description of these states, one has to exploit and justify various 
approximations. For instance, when electron binding energies exceed the spacings between rotational 
levels of the molecule, it is safe to neglect the corresponding non-Born-Oppenheimer couplings that can 
induce electron ejection, see Ref.~\cite{Crawford1977}. Likewise, when the binding energies exceed the 
vibrational level spacing, it is usually safe to neglect the analogous couplings 
that can lead to electron loss. 

\paragraph{Quadrupole-bound anions.}

The simplest model of a quadrupole-bound molecular anion considers the motion of an electron 
governed by the potential of a point quadrupole, 
$V_{\rm quad}(\vv{r}) = -(4\pi/5)(Q/r^3)Y_{20}(\theta,\phi)$, 
where $Q$ is the magnitude of the point quadrupole and $Y_{20}$ is a spherical harmonics. 
This potential can support bound states for any $Q>0$, i.e., unlike the case of a dipole, 
there is no critical value for the quadrupole moment to bind an electron. 
Similar to the binding by a point dipole, the electron is bound by a point quadrupole with an infinite 
energy due to the strong singularity of the potential at $r=0$. Therefore, additional assumptions about 
the interaction between the quadrupole molecule and the excess electron at small $r$ are invoked 
when the point quadrupole potential is considered as a source of binding. 
Such an approach was applied, for example in Ref.~\cite{Compton1996} to investigate 
the anionic state of the rotationally symmetric linear CS$_2$ molecule which does not possess 
a permanent dipole moment but has a sizable quadrupole moment $Q=3.3$~a.u. 
In order to model the Pauli repulsion between the attached electron and the core electrons of 
the molecule the point quadrupole potential was replaced by a positive infinite potential for values of 
$r$ smaller than some critical cut-off radius $R_c$. It has been found that for binding to occur, 
the cut-off radius must be small, $R_c \approx 0.3$~\AA, which is even much smaller than the C$-$S bond 
length ($\approx 1.56$~\AA) of the molecule. It is pointed out that this small value of $R_c$ does not 
imply that the overlap of the orbital of the attached electron with the orbitals of the core electrons of 
the molecule is large. Indeed, an analysis of the radial probability density of the attached electron 
presented in Ref.~\cite{Compton1996} shows that a sphere which contains $80$\% of the probability to 
find the electron has the radius $\approx 100$~\AA. 
This indicates a significant spatial extension of the state of the attached electron. 
 
Another example of a molecular anion which has been suggested 
to be quadrupole bound is the (BeO)$_2^-$ anion, see Refs.~\cite{Jordan1979, Gutowski1999}. 
In Ref.~\cite{Simons} it is mentioned that indeed for the latter anion the $d$-symmetry of the 
charge distribution of the excess electron is consistent with the angular part of the 
charge-quadrupole potential according to the spherical harmonics $Y_{20}$. 
However, it is pointed out that there are other interactions that contribute to the binding for this anion, 
e.g., valence short-range attractions in the regions of the two Be$^{+2}$ centres. The conclusion 
emerges that, even more than in the case of dipole-bound anions, it is impossible to find 
a species which can exclusively be called quadrupole-bound. 

According to accumulated experience, describing the dipole-bound and quadrupole-bound 
anions in terms of only long-range binding is nothing but a rather crude approximation. 
As counted in Ref.~\cite{Simons}, under this approximation relevant issues are ignored such as: 
(a) Coulomb and exchange interactions between the excess and inner-shell electrons,
(b) orthogonality of the extra electron's orbital to those of the other electrons in the molecule, and 
(c) indistinguishability of the electrons and thus the antisymmetry of the 
many-electron wave function within which the extra electron resides. 
Indeed, a picture where a single electrostatic potential guides an external electron becomes 
essentially inaccurate when this electron ``penetrates a neutral species''. 
In order to still consider the electronic attachment in the framework of a one-electron approach, 
some model modifications are usually introduced for the 
potentials~(\ref{charge-dipole})-(\ref{polarization}) at small distances. 
In addition, short-range terms are often applied to account for a valence-binding contribution. 
Examples of correspondingly modified potentials can be found, e.g., in Ref.~\cite{Desfrancois1996}. 
These potentials have been extensively applied to investigating the ground states of 
various molecular anions. 

\subsection{Combined-bound anions}

Combined binding mechanisms support certain {\it atomic anions}. These anions 
are expected to possess {\it resonance states} which correspond to high-spin configurations 
of the outer electronic shells. According to Refs.~\cite{Bunge1982, Nicolaodes1989}, 
the maximum spin polarized configurations are able to provide a strong unscreening of the nuclear 
charge, which is a necessary prerequisite to accommodate a further electron. 
Examples of such metastable states are the states $^3P^e~(2p^2)$ of the H$^-$ anion, 
$^4S^o~(2p^3)$ of the He$^-$ anion and $^4S^o~(1s\,2p^3)$ of the Li$^-$ anion. 

A combination of the short-range valence binding and binding due to long-range attractive forces 
can also be a mechanism providing {\it long-lived resonance states} of various {\it molecular anions}. 
Particularly interesting examples are the diatomic molecules N$_2$ and CO which are known 
as being unable to attach an extra electron in a conventional valence-bound stationary state. 

Experimentally, the anion N$_2^-$ has been observed in mass spectrometric investigations 
\cite{Gnaser1997, Middleton1999} suggesting a lifetime in excess of $10^{-5}$~s. 
The concept of high-spin states with several equivalent electrons has been applied in 
Ref.~\cite{Sommerfeld1998, N_2^-} to study the N$_2^-$ anion where three equivalent electrons are 
attached to an $N_2^{2+}$ core. In particular, in Ref.~\cite{N_2^-}, the state $^6\Sigma_u^+$ 
has been found to be stable against electronic autodetachment. 
In addition, its radiative decay to lower-lying electronic states 
can only occur via a spin-flip process, caused by e.g. spin-orbit interaction which is very weak. 
Hence this $^6\Sigma_u^+$ Feshbach resonance state should have a long lifetime providing the 
stability of the anion N$_2^-$ on the time scale of an experiment in a mass spectrometer. 

The molecules CO and N$_2$ are isoelectronic and consequently the anion CO$^-$ exhibits many 
similarities with the anion N$_2^-$. Subsequent theoretical investigations on excited states of 
this anion have been performed in Ref.~\cite{CO^-}. The same idea with respect to the construction 
of the anionic state has been exploited. An extra electron is valence-bound by a carbon atom occupying, 
together with the two outer electrons, the $2p$ shell. The resulting carbon anion, 
at reasonably large distances from the oxygen atom, can be considered as a point charge which 
polarizes the atom. This provides a long-range interaction between these species contributed by 
the charge-quadrupole, charge-induced dipole and charge-induced quadrupole attractions. 
Such an asymptotic behavior of the corresponding potential curve has been successfully confirmed by 
detailed {\it ab initio} calculations of the $^6\Pi$ high-spin sextet state of the anion CO$^-$ 
reported in Ref.~\cite{CO^-}. This state, similar to the $^6\Sigma_u^+$ state of the anion N$_2^-$, 
has been found to be a long-lived resonance state. 

Other examples of long lived resonance states of similar nature as the above ones are the high-spin 
states of the diatomic anion BF$^-$ \cite{BF^-} and the triatomic anion CO$_2^-$ \cite{CO_2^-}. 
The anion He$_2^-$, which has been discovered experimentally \cite{Bae1984}, can also be 
considered as supported by a combination of valence and long-range binding forces. A long-lived state 
of this anion has been identified theoretically in Refs.~\cite{Adamowicz1991a, Adamowicz1991b} as a 
$(1\sigma_g^2 \,\sigma_u\,2\sigma_g\,1\pi_u)~^4\Pi_g$ state, where two electrons are attached 
to a He$_2^+$ core. 

An interesting feature of diatomics which can form the high-spin anions has been discussed 
in Ref.~\cite{Kalcher}. These species can be expected to bind an extra electron in different states 
by either valence binding or  non-valence binding mechanisms. For ground states of various diatomic 
molecules (e.g., CMg, SiMg and others) the electric dipole moment exceeds the critical value 
of $1.625$~D which makes them candidates that possess dipole-bound negative ion states. 
One can also infer that all excited neutral states are accompanied by their respective 
dipole-bound anionic states. Therefore these species might not only support various valence-bound 
states, but additionally a number of dipole-bound anionic states, see Ref.~\cite{Kalcher2000}. 

The combined binding mechanism can also be revealed in some more complex species, for which it 
can support even stable {\it ground state} anions. These anions are formed by attaching an extra 
electron to molecules in a {\it Rydberg orbital}. A well-known example of a neutral Rydberg 
molecule is the NH$_4$ radical with an excited electron residing in a diffuse $s$-orbital of the 
core cation NH$_4^+$. Another example is the H$_3$C$-$NH$_3$ molecule 
in which an excited electron occupies the orbital centered on the 
closed-shell cationic site NH$_3^+$. For both examples the Rydberg orbitals have most of theirs 
densities outside the region where the cation's valence orbitals reside and can accommodate another 
electron. In this way the corresponding {\it double Rydberg anions} are formed. 
The attached electron becomes valence bound in the Rydberg orbital, while this orbital is linked to 
the cationic core or site of the anion by the long-range Coulomb attraction. 
The valence character of binding 
the extra electron in the Rydberg orbital is stressed in Ref.~\cite{Simons}, where it is pointed out 
that a successful treatment of such anions must allow for a correlated multiconfigurational 
treatment of at least the two electrons in the Rydberg orbital. The spatial properties 
of the Rydberg orbitals were investigated in Ref.~\cite{Cardy1986}, where the radial probability density 
of the highest occupied molecular orbitals (HOMOs) were presented for the tetrahedral structures of 
NH$_4$ and NH$_4^-$. Interestingly, the position of the maximum in the radial probability density is not 
substantially different for these species. In the direction of the NH bond this position is 
$2.19$~\AA~for NH$_4$ and $2.26$~\AA~for NH$_4^-$, while in the opposite direction it is 
$2.47$~\AA~and $2.61$~\AA, respectively. However, the HOMO of the anion is much more diffuse than 
that of the radical, which is a consequence of the repulsion between the two Rydberg electrons of 
the anion. This is illustrated by an increase of the radius of the sphere containing $90$\% of the 
probability of finding the electron: $4.52$~\AA~for NH$_4$ and $5.64$~\AA~for NH$_4^-$ in 
the NH bond direction, $4.73$~\AA~for NH$_4$ and $5.93$~\AA~for NH$_4^-$ in the opposite direction. 
The spatial extension of the Rydberg orbitals exceeds essentially the equilibrium distance between the 
atoms N and H in the radical ($1.050$~\AA) and in the anion ($1.0525$~\AA).

\section{Anions in the presence of a magnetic field}

The effects of an external magnetic field on the electronic structure of atomic and molecular 
systems have been the focus of theoretical investigations in recent decades. It is however remarkable that 
the corresponding investigations on anions basically concern the {\it valence-bound states} 
of a single anion, the negative hydrogen ion H$^-$. To our best knowledge, with the exception of 
Ref.~\cite{Guan2001} where strong magnetic field effects on the helium negative 
ion are investigated, there exist no detailed studies of other than H$^-$ anions in magnetic fields. 
Also, the impact of the magnetic field on the above-discussed non-valence-bound anionic states has 
not been discussed in the literature. 

The influence of the external magnetic field on the binding properties of the anion H$^-$ 
has been investigated for a broad range of field strengths, from zero up to astrophysically 
relevant values. The corresponding works can be presented by 
Refs.~\cite{Henry1974, Surmelian1974, Larsen, Park1984, Vincke1989, Larsen1, B_doubly_excited_H^-}, and the 
recent Ref.~\cite{Strong_B_H^-}. 
Already in early theoretical investigations, see, e.g., Refs.~\cite{Henry1974, Surmelian1974}, 
it has been demonstrated that exposing the anion H$^-$ to a magnetic field yields several 
interesting effects. The most striking is the appearance of {\it excited valence-bound states}. 
We remind the reader, that in field-free space, the negative hydrogen ion possesses a single bound state. 
Another interesting outcome is the fact that, with increasing field strength, the energetically lowest 
state of H$^-$ changes from the singlet, even-parity state, 
to the triplet, odd-parity state. A crossover of the corresponding energy levels 
has been found in Ref.~\cite{Surmelian1974} to occur at $B \approx 1.3\times10^4$~T. 
The situation is demonstrated in Figure~1, where we present the anionic total energies calculated  
in Ref.~\cite{Strong_B_H^-}, as functions of the field strength. The energy levels represent the lowest 
singlet and triplet states corresponding to the projections of the total electronic spin onto the direction 
of the field $S_z=0$ and $S_z=-1$, respectively. Also shown are the anionic detachment thresholds 
for these states. These thresholds are the total energies of the neutral atom and a free electron, both 
in the magnetic field, with the longitudinal component of the total electronic spin being preserved. 
The difference between the singlet and triplet threshold energies is gust given by the value of the 
magnetic field strength $B$ in atomic units ($1$~a.u. of the field strength corresponds to 
$2.3554\times10^5$~T). It features the well-known 
coupling of the energy of a system of ``bare'' electronic spins to the magnetic field strength, 
$E_{\rm spin} = BS_z$, see, e.g., Ref.~\cite{Landau1965}. 

Figure~2 shows the one-electron binding energies of the singlet and triplet states of the 
anion H$^-$ for a broad range of the magnetic field strengths. Dots, connected by smooth solid lines 
for convenience, represent the results of benchmark calculations from Ref.~\cite{Strong_B_H^-}. 
Both binding energies are determined with respect to the corresponding thresholds, i.e., 
they are the energies required to remove one electron from the anion without changing the 
value of $S_z$. 

The true physical detachment edge of the anion is given by the lowest of the detachment threshold energies 
which is the triplet detachment threshold energy shown in Figure~1. 
A relevant quantity is therefore the difference $I$ between this edge and the ground-state 
anionic energy. This energy changes, however, from the energy of the singlet state to the energy of the 
triplet state as the field strength is increased (see Figure~1). Such a quantity $I$ characterizes the true 
binding properties of the anion, 
regardless of the  spin orientations of the electrons before and after detachment. Notice that in reality 
the spin configuration of the anionic electrons is weakly perturbed by spin-orbit interaction. 
The value of $I$ can also be referred to as the true electron affinity of the hydrogen atom 
in the presence of a magnetic field. Remarkably, the earlier analysis of this affinity in 
Refs.~\cite{Henry1974, Surmelian1974} resulted in the striking conclusion that it takes negative values 
for $1.2\times10^3 < B < 3.3\times10^3$~T which means that the H$^-$ anion is unbound for 
that intermediate range of field strengths. In  contrast, the recent results 
obtained in Ref.~\cite{Strong_B_H^-} and presented in Figure~1 demonstrate that the 
lowest anionic energy level is below the triplet detachment threshold energy for any 
value of the field strength. In order to present in this review a plot of the true electron affinity 
as a function of $B$ we have used the reference data from Ref.~\cite{Strong_B_H^-} 
for interpolating the singlet and triplet binding energies, $I_s(B)$ and $I_t(B)$, respectively, 
by spline functions. The latter functions have been used to obtain a smooth curve $I(B)$ accordingly 
to the following expression: 
\be
I(B) = {\rm max}\{I_s(B)-B, I_t(B)\}~.
\label{IB}
\ee
This curve is given in Figure~3 by the solid line. For comparison, 
we also show by dashed lines the positive values of the electron affinity previously 
obtained in Ref.~\cite{Surmelian1974}. A gap between the two branches of the latter indicates 
the range of the field strengths where the anion H$^-$ was predicted to be not stable. 
The results from Ref.~\cite{Strong_B_H^-}, however, can be assumed to be more reliable. 
They clearly demonstrate that $I(B)>0$ at any $B$, i.e., the anion is always bound. 
With increasing $B$ the electronic affinity first decreases reaching a minimum value 
$I \approx 0.307\times10^{-3}$~a.u., and then increases again. The local minimum corresponds to a 
crossover of the singlet and triplet energy levels seen in Figure~1. 
The branch of the curve $I(B)$ with decreasing slope shows the energy required to remove one electron from 
the singlet bound state accompanied by the spin flip of the detaching electron at a magnetic field $B$. 
The slope of this curve is in rather good agreement with the results from Ref.~\cite{Surmelian1974}. 
The branch of the curve $I(B)$ with increasing slope represents the energy required to remove one electron, 
preserving its spin orientation, from the triplet state of the anion. Here, the previously obtained 
results from Ref.~\cite{Surmelian1974} deviate significantly from the correct dependence of $I(B)$. 
It is to be concluded that earlier variational calculations 
presented in Refs.~\cite{Henry1974, Surmelian1974} were not accurate enough for the triplet 
bound state of the anion H$^-$ leading to the wrong conclusion that the anion is unbound 
in the intermediate field regime. 

\section{Magnetically induced anions} 

This section, which is the main part of the present review, concerns the anionic states which 
exist {\it exclusively} due to the presence of an external magnetic field. In the 
preceding section, it was mentioned that the theoretical investigations of the binding properties 
of the negative hydrogen ion revealed the appearance of its excited valence-bound states 
when the anion is exposed to the magnetic field. Such a response of this specific anion 
to external magnetic field is already an interesting phenomenon. However, this is only one of many 
intriguing properties that one can expect when studying {\it any} anion in a magnetic field. 
The latter expectation is based on the formal mathematical proof provided in Ref.~\cite{Avron} 
that any singly charged negative ion possesses an {\it infinite number} of bound states in a magnetic 
field of {\it arbitrary strength}. Before, the role of the magnetic field in supporting the infinite 
manifold of anionic bound states was discussed in Ref.~\cite{AHS}. 

The property of the negative ions established in Ref.~\cite{Avron} means that they exhibit an unusual, 
essentially non-perturbative, response to external magnetic fields. Indeed, while in field-free 
space atomic and small molecular anions typically possess only one bound state and some even none, 
the number of bound states is infinite in a magnetic field of arbitrary strength. The fact that any 
neutral system can bind an electron in a magnetic field of arbitrary strength is very surprising and 
has far reaching experimental implications. However, the statement about the infinite manifold of the 
anionic bound states was formulated in Ref.~\cite{Avron} as the conclusion of a formal mathematical 
consideration of the properties of the underlying quantum operators. This investigation did not 
provide a transparent physical picture of the appearance of the infinite sequence of bound states. 
It also did not explicitly evaluate the corresponding anionic binding energies. Although a wide set 
of works focused on the influence of the magnetic field on low-lying anionic states 
(see the preceding section), surprisingly, for almost two decades there was no systematic treatment of 
the highly excited anions predicted in Refs.~\cite{AHS, Avron}. A physics based approach describing 
the infinite sequence of bound anionic states has been developed in Ref.~\cite{BSC2000}. 
The underlying binding mechanism can be regarded as a {\it non-valence binding} mechanism. 
The corresponding bound states are formed by the combined action of a polarization potential and the 
external magnetic field on the excess electron. In order to stress the exclusive role of the magnetic 
field for the binding of the electron these states were called in Ref.~\cite{BSC2000} 
{\it magnetically induced states}. Their binding energies have been explicitly evaluated. 
In Section~IV.A below we discuss the appearance of the infinite manifold of these states in more 
detail and present estimates of the corresponding binding energies. 

Magnetically induced states of negative ions exposed to external magnetic fields turn out to be 
of even more delicate nature than the theoretical considerations in Refs.~\cite{AHS, Avron, BSC2000} 
implied. These considerations assumed, as it is often done when calculating the electronic structure of 
atomic and molecular systems, that the anion is {\it infinitely heavy}, i.e. {\it spatially fixed}. 
However, this neglects, in the presence of the magnetic field, a coupling between the collective and 
internal degrees of freedom of the anions. A closer look at the problem, see, e.g., Ref.~\cite{BCS2001}, 
revealed, that the latter coupling is of major relevance for the binding mechanism supporting the 
magnetically induced anions. In Ref.~\cite{BSC2000} it has been already mentioned that one should expect 
the effects due to the finite anionic mass to be important for the properties of the magnetically 
induced states. This expectation is based on the theoretical experience accumulated within several 
decades separating the publications \cite{AHS, Avron} and \cite{BSC2000}. 
The important issue has been demonstrated that, in a magnetic field, the centre of mass (CM) motion does not 
decouple from the electronic motion and particular care must be taken when neglecting the coupling, see, 
e.g., Refs.~\cite{Johnson1983, Schmelcher1991, Schmelcher1994, Schmelcher1997, Schmelcher2000}. 
An intricate interaction between the CM and the internal degrees 
of freedom has been shown to lead to a variety of dynamical phenomena. For the neutral hydrogen atom, 
in Ref.~\cite{Schmelcher1992} it has been found that the transition from regularity to chaos in the 
classical internal motion is accompanied by a transition from quasi-periodic oscillations to a 
diffusional motion in the CM. For the He$^+$ ion, a permanent exchange of energy between the CM and 
electronic degrees of freedom can result in a dynamical self-ionization effect 
\cite{Schmelcher1995, Melezhik2000}. Rigorous quantum studies revealed essential influence of the CM 
motion in magnetic fields on the internal structure of the neutral hydrogen atom 
\cite{PavVer80, WRH81, ViBa88, PavMes93, ViDuBa92, Pot94} as well as of the positive hydrogen-like ions 
\cite{BaVi86, BezPavVen98}. The above cited results indicate that the properties of the magnetically 
induced anionic states could be strongly influenced by the coupling between the CM and internal motions. 
In view of this coupling the very statement that there is an infinite manifold of anionic bound states 
in the presence of a magnetic field becomes questionable. 

Accounting for CM effects requires the development of new theoretical approaches to describe 
{\it moving} magnetically induced anions. A meaningful approach has been worked out in 
Ref.~\cite{paper1}. It resulted in deriving a reduced-dimensionality Hamiltonian 
which describes the magnetically induced states of the anions of finite mass. 
The latter Hamiltonian accounts for the relevant coupling of the motion of the 
external electron to the motion of the neutral anionic counterpart (an atom or a small molecule).
The basic ideas of the underlying approach as well as its connection to the previously investigated 
model of an infinitely heavy anion are discussed in Section~IV.B below.

The Hamiltonian obtained in Ref.~\cite{paper1} has been applied in Refs.~\cite{BCS2001, paper2} to 
investigate the classical dynamics of excited anions in magnetic fields. These investigations 
demonstrate that the motion of the excess electron can change from a bound motion to an unbound one 
when the effects due to the CM motion are included. This proved that the coupling between 
the CM and internal motions is crucial for the stability of the magnetically induced states. 
The relevant classical dynamics is discussed in more details in Section~IV.C below.

Let us summarize up to this point. Following the formal proof given in Ref.~\cite{Avron} the 
theoretical investigations \cite{BSC2000} illuminate in physical terms the reasons why the number of 
bound states of anions in the presence of a magnetic field is infinite assuming the anions to be 
infinitely heavy. The corresponding binding energies have been explicitly evaluated. 
On the other hand, the approximation of infinite anionic mass was shown in Refs.~\cite{BCS2001, paper2} 
to be inadequate. Investigating the underlying classical dynamics of the moving anions in magnetic 
fields it was demonstrated that motional effects are indispensable for a complete and 
correct description of the dynamical processes as well as the structure of the anions. 
The latter investigations also mean that the coupling between the motions of 
the external electron and the neutral anionic counterpart can be expected to lead to a 
variety of interesting phenomena. 
For example, a quantized motion of the anion as a whole across the magnetic field 
can manifest itself in a discrete spectrum of the anionic bound states. 
Also, some magnetically induced states can become resonances when allowing the 
anion to move. Calculating the binding energies, positions and widths of resonances 
is, therefore, of major relevance when studying the properties of anions in 
magnetic fields. Finally, the number of magnetically induced bound states is by itself 
a quantity to be investigated taking account of the motional effects. 
To address the above questions, the quantum treatment of magnetically 
induced states for moving anions had to be further developed. Recently \cite{BC2003}, 
possibilities to quantise the Hamiltonian derived in Ref.~\cite{paper1} have been suggested. 
They are discussed in Section~IV.D below.

\subsection{Static anions: infinite manifold of magnetically induced bound states} 

The assumption that the anion is static (infinitely heavy) reduces the problem of its binding 
to the investigation of the electronic motion only. Such an approach has been exploited in 
Refs.~\cite{AHS, Avron, BSC2000}. As discussed above, Ref.~\cite{Avron} provides a very formal pure 
mathematical consideration employing advanced techniques from fundamental analysis. 
In Ref.~\cite{BSC2000}, several transparent physical approximations have been applied and justified 
in the course of the estimation of  the anionic binding energies. To begin with, one can expect the 
magnetically induced states to be loosely bound and spatially very extended states. 
The excess electron differs essentially in properties and behavior from the electrons of the neutral 
species which we will refer to as core electrons. One can therefore neglect the short-range quantum 
interactions (exchange etc.) of the excess electron with the core electrons. 
As a next step, the extra electron can be assumed to move much slower than the core ones. Then, one 
can exploit a quasistatic approximation neglecting non-adiabatic couplings between the states 
of the core electrons due to the motion of the external electron. Under the latter approximation, 
the binding mechanism can be treated as a {\it one-particle problem} where the motion of the 
excess electron is guided by an attractive potential $V(\vv{r})$ exerted by the 
neutral species. Those anions have been addressed whose neutral counterparts do not possess 
a permanent dipole moment. For them, the main source of the electrostatic attraction of the 
excess electron is a potential, polarization-like at large distances from the neutral species. 
In field-free space, the latter long-range attraction is not sufficient to bind the electron 
since it cannot support a confined electronic motion in three dimensions. The situation changes 
drastically if an external magnetic field is present. It confines the electronic motion in the plane 
perpendicular to the field, i.e., in two dimensions. In order to support a bound state, therefore, 
it is sufficient if the polarization potential can prevent the electronic escape along the field. 
As a consequence, the problem of the electronic binding can be considered as a 
{\it one-dimensional problem}. A further relevant approximation allows a description of the transverse 
motion of the excess electron in terms of the Landau orbitals indicating that this motion is 
dominated by the field. More explicitly, the part of the electronic wave function depending on the 
transverse coordinates $\vv{r}_\perp=(x,y)$ can be supposed to coincide with that for a free electron 
in a magnetic field, $\langle \vv{r}_\perp | n,s \rangle$. 

The above discussed steps are the essence of the procedure exploited in Ref.~\cite{BSC2000} to arrive 
at a one-dimensional Schr\"odinger equation describing the motion of the attached electron along 
the magnetic field. Within such an approach, the kinetic energy associated with the transverse 
motion of the attaching electron can take the discrete values
\be
\eL_n = B \left( n + \frac{1}{2} \right)~, \;\;\; n = 0,1,2,\ldots~,
\label{eL}
\ee
which are nothing but the Landau energies of a free electron. The quantum number $n$ enumerates the 
electronic Landau levels. The electronic longitudinal angular momentum can equal $l_z = -s$ 
where $s = -n, -n+1, -n+2, \ldots$. The quantity $\eL_0 = B/2$ determines the continuum threshold for 
the states of the external electron. Typically, the states that are truly bound relate to the ground 
Landau manifold, $n=0$. The electron can therefore be attached to the neutral species with 
the values of $l_z$ corresponding to $s=0,1,2,\ldots$. 

It is important that for the static anion $l_z$ represents an {\it integral of motion}. 
Hence the quantum number $s$ can be conveniently used to label the states of the 
attaching electron. For each $s$, one can speak of a specific one-dimensional potential, 
$V_s(z) = \langle 0,s | V(\vv{r}) | 0,s \rangle$, which enters the corresponding 
Schr\"odinger equation describing the electronic motion along the magnetic field. 
The latter potential can bind the electron in at least one quantum state. Subsequently, 
a negative ion can be formed in an infinite manifold of the magnetically induced states. 
Assuming the polarization character (\ref{polarization}) for the potential $V(\vv{r})$, 
the corresponding binding energies have been explicitly evaluated in Ref.~\cite{BSC2000} 
using the weak-coupling one-dimensional theory: 
\begin{eqnarray}
\E_0 &=& 0.31\kappa^2 B^2~,\;\;\;\;\;\;s=0~,
\label{binding_0} 
\\
\E_s &=& 0.12\kappa^2 B^3 \delta_s^2~,\;\;\;s=1,2,\ldots~,
\label{binding_s}
\\
\delta_1 &=& 1~,\;\;\;\delta_s = [1-(1.5/s)] \delta_{s-1}~,
\nonumber 
\end{eqnarray}
where $\kappa$ is the scalar polarizability of the neutral species and $B$ is the magnetic field strength, 
both in atomic units. By specifying the polarizability $\kappa$, the above estimates can be applied 
to any atomic or small molecular negative ion assuming a vanishing permanent electric dipole moment. 
These estimates establish the dependence of the binding energies on the magnetic field strength and 
on the quantum number $s$: for $s=0$ the binding energy scales according to $B^2$, while for 
$s=1,2,\dots$ they scales according to $B^3$; for large $s$ the binding energies behave as 
$-\E_s \propto s^{-3}$. In Ref.~\cite{BSC2000} it has also been verified that the estimates 
(\ref{binding_0}) and (\ref{binding_s}) are in complete agreement with results from a numerical 
solution of the Schr\"odinger equation for the motion of the external electron along the magnetic field. 

Let us point out a few relevant aspects. The estimates (\ref{binding_s}) for the binding energies of the 
``excited'', i.e. corresponding to $s=1,2,3\ldots$, magnetically induced states are obtained 
assuming that there is no deviation of the attractive potential $V(r)$ from the polarization 
one (\ref{polarization}) at small $r$. In the transverse plane, the electronic density distributions 
corresponding to the Landau orbitals with $s>0$ are progressively shifted from the origin with 
increasing $s$. As a result, the short-range behavior of the potential $V(r)$ plays no role in affecting 
the motion of the excess electron. In other words, the magnetically induced states with $s>0$ 
are supported exclusively by the long-range polarization attraction of the electron to the neutral 
species. For the ``ground'', i.e. corresponding to $s=0$, magnetically induced state, the short-range 
behavior of the potential $V(r)$ is important. One has to model this potential at small $r$ differing 
it from the polarization one. This can be achieved, e.g., by introducing a cut-off of the polarization 
potential (\ref{polarization}) at distances smaller than a typical size of the neutral species. 
For the anion H$^-$, a model potential has been used in Ref.~\cite{BSC2000} related to the known 
energy \cite{Wallis1960} of an electron moving in the Coulomb fields of the two static charges: 
of the nucleus (proton) and of the external electron. The numerical coefficient in the estimate 
(\ref{binding_0}) is therefore model-dependent. The related feature is also the scaling behavior of this 
binding energy with respect to the magnetic field strength different from that of the binding energies 
for $s>0$. However, as discussed in Refs.~\cite{BSC2000, paper1}, there are significant reasons to believe 
that the scaling behavior of $\E_0$ with respect to both the polarizability and the magnetic field 
strength has a universal character, while the dependence of the numerical coefficient on the short-range 
behavior of the potential $V(r)$ is rather weak. 

Another remark is appropriate. The estimate (\ref{binding_0}) should be applied with particular care 
to those anions which possess ground bound states in field-free space. Indeed, the wave function 
describing the motion of the excess electron for the anions in field-free space exhibits normally 
the same nodal structure and parities as the wave function of the magnetically induced $s=0$ state. 
The magnetically induced $s=0$ state should therefore strongly mix with the conventional anionic state 
in case the latter exists. Such a mixture can be expected to change the properties and in particular 
to shift the energies of the corresponding states. 

The above comments on a character of binding of the excess electron in the $s=0$ magnetically 
induced state provide an analogy of this state with the non-valence bound states of the anions 
in field-free space. As pointed out in Section~II.B, for the latter states the short-range valence 
forces play also a significant role. However, if in field-free space one can hardly find an example of 
anionic state supported exclusively by a long-range interaction, such states appear in the presence of 
a magnetic field. A detailed physics-guided investigation~\cite{BSC2000} demonstrates that 
the magnetically induced states $s=1,2,\ldots$ are exclusively {\it non-valence bound}. This investigation 
confirms the general conclusion previously obtained in the mathematical study \cite{Avron}: for static 
anions, the number of the bound states is infinite in the presence of a magnetic field. 
For those neutral species which cannot bind an extra electron in field-free space, the infinite 
sequence of the magnetically induced states also includes the $s=0$ state, while for the anions 
possessing a bound (ground) state in field free-space an accurate description of the ``ground'' 
magnetically induced state might go beyond the approach developed in Ref.~\cite{BSC2000}. 

Table~1 presents the binding energies of the magnetically induced states exclusively for the 
group of atomic anions which do not exist in field-free space (see Section~II.A). 
For the latter anions, one can confidentially apply the estimate (\ref{binding_0}) for the lowest 
magnetically induced states corresponding to $s=0$. Two values of the magnetic field strengths, 
$B=10$~T and $B=30$~T, respectively, are chosen which can be achieved at laboratories. 
For each value, the binding energies for the states $s=0$ and $s=1$ are indicated. Also indicated 
are the corresponding atomic polarizabilities according to Ref.~\cite{Miller1996}. Among them the 
smallest polarizability is that of the He atom, while the biggest one is that of the Cd atom. 
Correspondingly, at $B=30$~T, the He$^-$ anion has the smallest binding energies, 
$\E_0 \approx 2.63\times10^{-4}$~meV and $\E_1 \approx 1.3\times10^{-8}$~meV of the ground and 
first excited magnetically induced states, respectively. In contrast, the anion Cd$^-$, 
at the same magnetic field strength, exhibits significantly larger binding energies of the 
corresponding states, $\E_0 \approx 0.32$~meV and $\E_1 \approx 1.6\times10^{-5}$~meV, respectively.  

Among all atoms, the atom Cs possess the largest polarizability, $\kappa \approx 403$. Small molecules 
Rb$_2$ and Cs$_2$ possess however even bigger polarizabilities \cite{Miller1996}, 
$\kappa \approx 533$ and $\kappa \approx 702$, respectively. At a magnetic field of $10$~T, 
the estimates (\ref{binding_s}) provide for the corresponding magnetically induced excited 
anionic states the binding energies $\E_1 \approx 4\times10^{-5}$~meV, 
$\E_1 \approx 7\times10^{-5}$~meV and $\E_1 \approx 1.2\times10^{-4}$~meV, respectively. 

The approach developed in Ref.~\cite{BSC2000} can also describe magnetically 
induced states of exotic ``muonic'' anions formed by attaching a muon to a neutral species. 
For such anions, there is no exchange interaction between the attaching muon and the electrons 
of the neutral anionic counterpart, so there is even no need to invoke the approximation 
which neglects this interaction. The final estimates of the binding energies exceed these given 
by Eqs.~(\ref{binding_0})-(\ref{binding_s}) by the ratio of the muonic mass to the electronic mass, 
$\me/m_\mu \approx 207$. 

\subsection{Beyond the approximation of spatially fixed (static) anions}

In order to describe the magnetically induced anionic states taking into account the coupling between 
the CM and the excess electron, a corresponding approach was developed in Ref.~\cite{paper1} that 
allows a reduction of the dimensionality of the problem. In the following we sketch the main steps of 
derivation of the corresponding Hamiltonian. 

The same idea as for the case of the static anion has been exploited that the extra electron is loosely 
bound and possesses a low velocity compared to that of the tightly bound electrons of the neutral species. 
Suppose for simplicity that the anion is formed by attaching the extra electron to an atom. 
In a first step, the Hamiltonian of the system is expressed in terms of the CM coordinate $\vv{R}$ of 
the neutral atom and a relative coordinate $\vv{r}$ of the extra electron with respect to the atomic CM. 
The relative coordinates of the core electrons are, however, specified with respect to the nucleus. 
Then a unitary gauge-like transformation allows one to eliminate coupling terms of the 
motion of the core electrons to both the motions of the atomic CM and of the excess electron 
which involve the canonical momenta of the core electrons. After that the Hamiltonian can be 
averaged over the motion of the core electrons using the procedure exploited in Ref.~\cite{BSC2000}. 
One obtains 
\begin{eqnarray}
H &=& \frac{\vv{\pi}^2}{2\mat} + \He + V(r)~,
\label{H_av}
\\
\vv{\pi} &=& \vv{P} + \frac{1}{2}\,\vv{B}\vtimes\vv{R} 
           - \vv{p} + \frac{1}{2}\,\vv{B}\vtimes\vv{r}~,
\label{pi}
\\
\He &=& \frac{1}{2} \left( \vv{p} + \frac{1}{2}\,\vv{B}\vtimes\vv{r} \right)^2~.
\label{He}
\end{eqnarray}
The Hamiltonian (\ref{H_av}) allows a description of the moving negative ion as a neutral atom 
interacting with an external electron, i.e., it involves {\it six degrees of freedom}. 
The first term in (\ref{H_av}) represents 
the kinetic energy of the atom and the coupling of the atomic CM to the electron, 
$\mat$ is the mass of the atom. $\vv{P}$ and $\vv{p}$ are the momenta conjugated 
to $\vv{R}$ and $\vv{r}$, respectively. $\He$ describes a free electron in a magnetic 
field and $V(r)$ is an attractive potential which links the electron to the atom. 
When $r$ exceeds the atomic size, this potential takes on the appearance of a 
polarization potential: $V=-\kappa/(2r^4)$. It should be emphasized that the Hamiltonian 
(\ref{H_av}) involves the parameters which characterise the atom as a whole i.e. 
the atomic mass $\mat$ and polarizability $\kappa$. Therefore, it can be likewise applied to 
study magnetically induced bound states of any anion, by specifying these parameters for 
the underlying neutral system (atom or small molecule). For a non-spherical molecule $\kappa$ 
may depend on the orientation of this molecule with respect to the magnetic field axis. This can 
be included in Eq.~(\ref{H_av}) by defining $V=V(\vv{r})$ appropriately.

For further treatment of the Hamiltonian (\ref{H_av}) it is appealing to introduce new canonical variables 
different from the Cartesian ones. In a magnetic field, a free charged particle performs a Larmor rotation 
around its guiding centre in the plane perpendicular to the field, see, e.g., Ref.~\cite{Johnson1983}. 
A bound anion as an entity is expected to exhibit a Larmor-like rotation. Having this 
in mind, the motion of the neutral counterpart can also be conveniently described in terms of the 
variables related to the Larmor rotation. Instead of the atomic transverse coordinates and momenta 
one can introduce two sets of coordinates: $\Xc$ and $\Yc$ which determine the location of the 
guiding centre, and $\XL$ and $\YL$ which provide the location of the atom with respect to this 
centre. Similar coordinates can be introduced for the excess electron: $\xc$ and $\yc$ describe the 
location of the electronic guiding centre with respect to the atom and $\xL$ and $\yL$ provide the position 
of the electron with respect to $(\xc,\yc)$. From the quantum point of view for free charged particles, 
the cylindrical radii corresponding to the above pairs can only take on the following discrete values, 
\begin{eqnarray}
\Rc^2 &=& \Xc^2+\Yc^2 = (2N_0+1)/B~, \;\;\; 
\RL^2 = \XL^2+\YL^2 = (2N+1)/B~, 
\nonumber \\
\rc^2 &=& \xc^2+\yc^2 = (2n_0+1)/B~, \;\;\;\;\;\;\; 
\rL^2 = \xL^2+\yL^2 = (2n+1)/B~, 
\label{quantum_values}
\end{eqnarray}
where $N_0$, $N$, $n_0$ and $n$ are non-negative integer numbers. It is worthy of notice that 
the kinetic energy of the electronic Larmor rotation is determined by the radius of the 
Larmor orbit and equals $0.5B\rL^2$. The last relation of Eqs.~(\ref{quantum_values}) naturally 
provides the latter energies to be the Landau energies (\ref{eL}). 

Although the reader might have an impression that the above quantities are introduced formally, 
they do relate to physically meaningful types of motion in magnetically induced anions. 
For an illustration of these quantities, Figure~4 shows the mutual arrangement of the origin and the 
guiding centres as well as the Larmor orbits for the atom and the external electron. 
To provide an idea of the separation between the atom and the electron, the displacements are chosen 
in Figure~4 according to Eqs.~(\ref{quantum_values}) for $N_0=2$, $N=0$, $n_0=1$ and $n=0$ for 
the magnetic field strength $B=10$~T. The choice of $N=n=0$ simulates the atom and the external 
electron moving on the ground Landau orbits. 

Having introduced the new sets of coordinates, it is possible to utilise the conservation of the 
total pseudo-momentum parallel and perpendicular to the magnetic field and to eliminate 
two atomic CM degrees of freedom from the Hamiltonian. 
Then the motion of the ion can be described in terms of {\it four degrees of freedom}. 
Three degrees of freedom are associated with the motion of the extra electron with respect 
to the atomic CM and relate to the pairs $\{\xc,\yc\}$, $\{\xL,\yL\}$ and $\{z,p_z\}$. 
One more degree of freedom corresponds to the atomic motion in terms of the pair $\{\XL,\YL\}$. 

Basic features of the anionic motion in different degrees of freedom are demonstrated in Figures~5-7. 
They show the classical trajectories for the anion Cs$^-$ in a magnetic field $B=40$~T. 
The initial conditions simulate a bound anion, occupying as an entity the ground Landau level, while 
the excess electron is attached to the atom in a state corresponding to the magnetically 
induced state $s=3$ of a static anion. We refer the reader to Ref.~\cite{paper1} for the details. 
Figure~5 shows the anionic motion in terms of the motions of the 
atom and the external electron. The slowest motion is exhibited by the atom and shown in Figure~5a. 
The motion of the external electron along the magnetic field exhibits faster 
oscillations (see Figure~5c). The electronic motion transverse to the magnetic field is 
a combination of a slow drift of its guiding centre $(\xc,\yc)$ 
and an extremely fast Larmor rotation around it (see Figure~5b). 
This demonstrates that indeed the pairs $\{\xc,\yc\}$ and $\{\xL,\yL\}$ naturally identify 
{\it slow} and {\it fast} degrees of freedom, respectively, for the motion of the external electron.

Figure~6 shows the electronic drift motion across the field in terms of the slow variables 
$\{\xc,\yc\}$. The electronic guiding centre follows a circular-like trajectory around the atom. 
The radius of this trajectory however varies, and the variation corresponds essentially to the 
modulation of the atomic trajectory (see Figure~5a). The separate Larmor motion of the electron is 
shown in Figure~7a. The number of rotations of the electron around its guiding centre corresponds to 
the number of the fast oscillations of the electronic Larmor radius involved in the figure. It is 
huge compared to the oscillations of all other slow degrees of freedom on this time scale. To be 
specific, for a time interval of $8$~ns the radius of the electronic guiding centre exhibits 
$1$ oscillation. For the same time interval, the number of the longitudinal oscillations of the electron 
is $84$ while the number of the Larmor rotations is $17906$. 
 
An important feature of the fast electronic motion is explicitly seen
in Figure~5b: the time evolution of the electronic Larmor radius does not exhibit
a slow modulation as the time evolution of the guiding centre does.
This indicates that the atomic motion to a good approximation does not interact
with the electronic Larmor rotation.
In other words, being initially placed on a Landau orbit,
the external electron always follows it with its Larmor radius
showing only tiny fluctuations around a constant mean value.
The Larmor radius can therefore be considered as an {\it approximate integral of motion}. 

The character of the anionic motion discussed above allows one to draw several important conclusions. 
The main effect of the interaction between the neutral species and the external electron for a moving 
anion is due to the coupling between the CM of the neutral species and the guiding centre of the 
external electron i.e. between the slow degrees of freedom for the anion. 
Similar to the case of the infinitely heavy anion, when describing the magnetically 
induced states of moving anions, a physically meaningful approximation may still assume a conservation, 
at the level of the zero-point Landau energy, of the kinetic energy of the electronic motion transverse 
to the field. However, in contrast to the static case, the electronic longitudinal angular momentum 
$l_z$ is no longer an integral of motion for a moving anion due to the drift motion of the 
guiding centre $(\xc,\yc)$. 

In order to average the Hamiltonian (\ref{H_av}) over the fast electronic Larmor rotation, a successful 
quantum procedure has been developed in Ref.~\cite{paper1}. When studying the electronic motion in 
terms of the variables discussed above (we refer also to Figure~4), one can introduce 
{\it one dimensional} states, $\langle \xL | n \rangle$ or $\langle \yL | n \rangle$, equivalently 
describing pure Larmor rotations. The latter states provide the electronic Landau energies (\ref{eL}) 
but they do not specify the electronic guiding centre or the angular momentum $l_z$. Applying the 
ground ($n=0$) states for the desirable averaging involves a series of non-trivial transformations. 
They result in the following effective {\it three-dimensional} Hamiltonian that applies to 
the magnetically induced states of moving anions: 
\begin{eqnarray}
\Heff &=& H_1 + H_2~,
\nonumber \\
H_1   &=& \frac{B^2}{2\mat}\,\left[ \left(\XL+\xc\right)^2 + \left(\YL+\yc\right)^2 \right]~,
\nonumber \\
H_2  &=& \frac{p_z^2}{2\mu} + \Veff(\rc^2,z^2)~.
\label{Heff}
\end{eqnarray}
The term $H_1$ represents the kinetic energy associated with the motion of the neutral species 
across the magnetic field. It also involves the coupling between the transverse degrees of freedom 
of this species and the extra electron. $H_2$ describes the motion of the extra electron relative to the 
neutral species, $\mu=\me\mat/(\me+\mat)$ is the reduced mass of this electron. 
The main ingredient of the latter Hamiltonian is an effective potential 
$\Veff(\rc^2,z^2) = \langle 0 | V(\vv{r}) | 0 \rangle$. It has been worked out in Ref.~\cite{paper1} 
in an explicit form, 
\begin{eqnarray}
&\,& \Veff(\rc^2,z^2) = - \frac{\kappa B^2}{8}\,
                          \int_0^\infty {\rm d}\xi\,\frac{\xi\exp(-v\xi)}{(1+\xi)^f}~,
\nonumber \\
&\,& v = 0.5 B z^2~,\;\;\;
     f = 0.5 \left( B\rc^2 + 1 \right)~.
\label{Veff}
\end{eqnarray}
The Hamiltonian (\ref{Heff}) possesses an integral of motion determined by the operator 
\be
\Lt = \frac{B}{2}\left(\RL^2-\rc^2\right)~.
\label{L}
\ee
In this context, i.e. having eliminated irrelevant degrees of freedom, it can be regarded as the 
total longitudinal angular momentum of the anion. As typical for many-body systems, the latter integral 
of motion can hardly be implemented to additionally reduce the dimensionality of the 
Hamiltonian~(\ref{Heff}) in the framework of further studies of the quantum properties. 

To make contact with the bound states of the infinitely heavy anion, we notice that in the limit 
$\mat\to\infty$ the electronic guiding centre displacement becomes an integral of motion. 
It can take the discrete values $\rc^2=(2s+1)/B$, $s=0,1,2,\ldots$ (cf. Eqs.~\ref{quantum_values}), 
which correspond to the electronic longitudinal angular momentum $l_z=-s$. For these fixed values 
of $\rc^2$, the effective potential (\ref{Veff}) has been shown in Ref.~\cite{paper1} to coincide 
with the one-dimensional potential $V_s(z)$ that determines the properties of the magnetically induced 
states for static anions (see Section~IV.A). This means that, at the selected values 
$\rc^2=(2s+1)/B$, the Hamiltonian $H_2$ has the bound eigenstates with the energies 
$-(\mu/\me)\E_s$, where $\E_s$ can be estimated according to Eqs.~(\ref{binding_0})-(\ref{binding_s}). 
The latter states, however, are no longer the eigenstates of the Hamiltonian (\ref{Heff}) for finite 
mass $\mat$ of the neutral species. Considering the complete problem they mix with the eigenstates 
of the Hamiltonian $H_1$ when the anion is allowed to move. 

Possibilities of a further quantum treatment of the Hamiltonian~(\ref{Heff}) will be considered 
at the end of this review. At this point we emphasize that, with the fastest degrees of freedom being 
removed in the course of the above sketched derivation, the numerical propagation of the classical 
motion for long times with very high stability is possible. Relevant investigations including 
statistical studies of the motion-induced detachment of anions are discussed 
in the following section. 

\subsection{Motion-induced dynamics and decay of the magnetically induced states}

In Ref.~\cite{paper2} the classical motion with respect to the degrees of freedom involved in the 
Hamiltonian~(\ref{Heff}) has been investigated. The classical trajectories have been calculated for 
fixed {\it integer} values of the total longitudinal angular momentum~(\ref{L}) i.e. according to the 
quantised system.

For classical simulations of the moving magnetically induced anions one has to establish 
relevant energy shells for which the trajectories propagate. 
Without having obtained the energy spectrum of the Hamiltonian (\ref{Heff}) 
from quantum considerations, some assumptions about relevant energy values are required. 
It is legitimate to assume as initial conditions such conditions which simulate the bound anion 
in the absence of the coupling between the motions of the neutral species and the extra electron. 
Starting from such conditions, the classical trajectories can be propagated taking into account 
the motional coupling thereby elucidating its role for the dynamics of the anion. 
Such an approach has been exploited in Ref.~\cite{paper2}. By the choice of initial conditions 
the anion was placed, as an entity, on a Landau orbit, with the extra electron bound 
to the neutral species corresponding to that for a static anion. The trajectories have then 
been propagated for the energies defined by two quantum numbers, 
\be
E_{Ns} = \frac{B}{M}\left(N+\frac{1}{2}\right) - \frac{\mu}{\me}\,\E_s~,
\;\;\; N = 0,1,2,\ldots~,
\;\;\; s = 0,1,2,\ldots~.
\label{Shell}
\ee
The first term in Eq.~(\ref{Shell}) is the Landau energy of the anion (cf. Eq.~(\ref{eL}) for 
the electronic Landau energies), $M=\mat+\me$ is the total mass of the anion, while $\E_s$ is 
the binding energy of the extra electron in a static anion given in 
Eqs.~(\ref{binding_0})-(\ref{binding_s}). 

Evidently, when the energy value $\Heff=E_{Ns}$ is negative, the classical motion is confined 
to a finite portion of phase space. If the energy is positive, one might expect 
to observe detaching trajectories corresponding to an infinite motion 
and phase space of the system. These features of the classical dynamics are demonstrated in 
Figures~8, 9 and 10 which show the motion of the Cs$^-$ ion in a magnetic field 
$B=10$~T. All the examples correspond to initial conditions according to the 
magnetically induced bound state $s=1$ of the static ion. In order to define the 
spatial location of the atom it was chosen to coincide with that presented in Figure~4.

Figure~8 shows the motion of the ion initially placed on the ground Landau level $N=0$, 
for the value $\Lt=0$. The corresponding anion energy is negative and the available phase space 
is finite providing a bound motion of the anion. This motion is comprised of the atomic rotation 
around the fixed guiding centre (Figure~8a), the rotation of the electronic guiding centre around 
the atom (Figure~8b) and the oscillations of the electronic longitudinal coordinate (Figure~8c). 
Figure~9 shows the atomic trajectories for a positive value of the Hamiltonian (\ref{Heff}) 
which corresponds to an initial excitation of the anion according to the Landau level $N=10$. 
The propagated trajectories relate to the different integer values of the integral of motion 
$\Lt$ which vary from $3$ to $18$. The smaller values of $\Lt$ correspond to larger 
initial curvature the atomic trajectory resulting in a more efficient dynamical energy 
exchange between the atomic and the electronic degrees of freedom. As a result, for 
$\Lt=3,4,5,6$ and $7$ the latter fluctuating energy exchange triggers the autodetachment process 
and the atomic motion becomes unbound. For $\Lt=8,9,\ldots,18$ these fluctuations are insufficient 
to cause the autodetachment and the anion moves as a bound system with the atom rotating on 
circular-like orbits. 

In Figure~10, the electronic motion corresponding to the atomic trajectories shown 
in Figure~9 is presented. The time variations of the electronic longitudinal coordinate $z$ 
and the transverse displacement of the guiding centre $\rc$ from the atom are given. 
The illustrations for $\Lt=5$ and $7$ (Figures~10a,b) correspond to the detaching atomic 
trajectories, and the trajectories for $\Lt=8$ and $10$ (Figures~10c,d) relate to the finite 
circular-like atomic motion. One immediately realises that for the detaching motion 
both the longitudinal and transverse separation between the atom and the electron 
increase unlimited with time. For the bound anion, both displacements oscillate with time, 
and the amplitude as well as the frequency of the fast oscillations of $z$ are modulated by the 
slower oscillations of $\rc$. The latter modulation decreases for increasing $\Lt$ 
indicating that the efficiency of the energy exchange between the atom and the electron reduces 
with the increase of the radii of the atomic orbit. It is also seen that the frequency of the 
longitudinal electronic oscillations is maximal while the corresponding amplitude is 
minimal for minimal transverse separation between the electronic guiding centre and atom.

In order to investigate the stability of the anions it is sufficient to consider only 
the relative motion of the external electron with respect to the atom. The corresponding 
classical equations of motion have been derived in Ref.~\cite{paper1} exploiting the conservation 
of the anionic angular momentum (\ref{L}). They couple {\it two degrees of freedom} associated 
with the pairs $\{\rc,\varphi\}$ and $\{z,p_z\}$ of conjugated variables. 
The angle $\varphi$ is the angle between the vectors $\vv{R}_{\rm L}=(\XL,\YL,0)$ and 
$\vv{r}_{\rm c}=(\xc,\yc,0)$. 

The reduction of the problem of anionic stability to a two-dimensional one is physically meaningful. 
The difference in the number of the relevant dimensions, i.e. between {\it one} and {\it two}, 
is the essence of the difference between electronic attachment for a {\it static} and 
{\it moving i.e. realistic} anion in a magnetic field. Indeed, the magnetic field confines the 
extra electron in two dimensions. 
Its binding to a static neutral species then becomes a subject of a one-dimensional motion. 
However, a neutral species of a finite mass is no longer a static centre. The motion of this species, 
since it is a neutral particle, is not confined by the magnetic field. Therefore, transverse separation 
between the neutral species and excess electron is an additional degree of freedom which plays a 
relevant role in the attachment and detachment processes. 

The two-dimensional analysis of the anionic stability has been exploited in Ref.~\cite{paper2} 
to study the decay of the initially bound anionic states via the motional-induced coupling. 
Classical trajectories have been propagated corresponding to different initial conditions 
associated with certain energy shell (\ref{Shell}). For each trajectory of such an ensemble 
the detachment time have been calculated being a measure of the time required to change the 
character of motion from initially bound to an unbound one. In this way the autodetachment curves 
have been obtained which show the fraction of detached trajectories depending on the propagation 
time. 

As a relevant example we show in Fig.~11 the detachment curve for the ion H$^-$ moving 
in a typical laboratory magnetic field of strength $B=1$~T. The lowest energy shell (\ref{Shell}) 
for $N=s=0$ has been selected which is positive, $E_{Ns}=1.046\times10^{-9}$~a.u., and therefore 
relates to autodetachment. The classical simulations show that the autodetachment process is essentially 
completed for the complete ensemble of trajectories within $12$~ns. This time scale is a measure 
for the lifetime of the anion H$^-$ prepared initially in the $N=s=0$ state. 
Also shown in Figure~11 are the typical detaching trajectories associated with the different 
parts of the autodetachment curve. They feature two essentially different scenarios of the electronic 
motion, associated with the two possible signs for the derivative $\dot{r}_{\rm c}$ at the initial time. 
For positive signs, the transverse separation of the electronic 
guiding centre from the atom starts to increase from the very beginning of the propagation of the 
trajectory. As a result, the effective potential weakens and already after one longitudinal oscillation 
the electron is no longer sufficiently attracted by the atom to support a bound motion (see the bottom 
insert in Figure~11). For negative signs, the transverse displacement $\rc$ initially decreases making 
the interaction $\Veff$ stronger. Consequently, the longitudinal electronic motion undergoes a series 
of oscillations which become increasingly denser with the decrease of $\rc$ as the upper insert in 
Figure~11 demonstrates. This part of the motion is associated on average with energy transfer from 
the electronic to the atomic degrees of freedom. At later times the opposite tendency is observed and 
the electron acquires energy from the atomic motion. Then $\rc$ starts to increase and the following 
motion finally becomes a detaching one. 

Another example is presented in Figure~12. It illustrates the autodetachment decay of the magnetically 
induced Cs$^-$ anion. This anion is much heavier than H$^-$ and therefore expected to be more stable with 
respect to the motion-induced coupling. Besides, for fixed nucleus, its magnetically induced states 
are more strongly bound, due to the large polarizability of the Cs atom. For the same magnetic field 
strength, $B=1$~T, the energy shell (\ref{Shell}) for the anion Cs$^-$ corresponding to $N=s=0$ is 
negative providing a stable bound motion. For $N=0$ and higher $s$ the energy shells are positive. 
The corresponding states undergo the autodetachment process, and the related curves are shown 
in Figure~12. They demonstrate an interesting feature that for higher internal excitations of the 
anion its motion-induced decay is much slower than for low excitations. For example, the typical time 
for the complete autodetachment decay of all anionic configurations $N=0$, $s=2$ is $3000$~ns. It is 
by a factor of two larger than the corresponding time of $1400$~ns for the decay of the anions initially 
prepared for $N=0$, $s=1$. The reason for the longer lifetimes of the higher internal excitations relates 
to the fact that the higher excitations correspond to larger initial transverse separations between the 
atom and the excess electron. This reduces the interaction and consequently the energy exchange between 
the particles. Extrapolating this classical feature to properties of quantum states of the magnetically 
induced anions one may expect that the energetically higher resonance states have longer lifetimes. 

The character of the autodetaching motion for the anion Cs$^-$ is illustrated in Figure~13 by 
three typical examples of the trajectories contributing to the three distinct parts of the anionic 
autodetachment curve for $s=3$ shown in Figure~12. The trajectories (a) and (b) correspond to a 
relatively fast autodetachment, when the transverse separation between the electronic guiding centre 
and the atom increases staring from the initial time. Interestingly the evolution $\rc=\rc(t)$ 
is the same for the trajectories (a) and (b) which differ significantly with respect to the variation 
of the electronic coordinate $z=z(t)$. This indicates that the transverse electronic degrees of freedom 
are influenced predominantly by the atomic motion and less by the electronic motion along the field. 
The trajectory (c) first experiences a reduction of the separation between the electronic guiding centre 
and the atom. The fastest longitudinal oscillations occur close to the minimal value for $\rc$. 
In total 189 oscillations occur with respect to $z(t)$ before the trajectory becomes detaching. 

An analysis of the dynamics of the magnetically induced anions reveals that the energy exchange between 
the neutral species and the excess electron mostly involves the transverse electronic degrees of freedom 
associated with the variation of the guiding centre displacement $\rc$. The magnetically induced states 
which are bound under the approximation of a static neutral species can decay when the latter is allowed 
to move. The channel of decay relates to increasing $\rc$ i.e. transverse separation between the particles. 
This weakens the effective two-dimensional potential $\Veff=\Veff(\rc^2,z^2)$ which, above some critical 
value of $\rc$, looses the ability to support bound electronic motion along the magnetic field. 

On the one hand, the classical simulations provide the reasons to assume that the inclusion of the 
collective motion terminates the infinite series of the magnetically induced bound states predicted 
in \cite{Avron}. On the other hand, they reveal the times of autodetachment to be long enough to 
speak of the corresponding states as of anionic resonances. Further quantum investigations are therefore 
of great relevance in order to quantitatively describe properties of bound and resonance states for
moving magnetically induced anions. 

It is naturally desirable to determine the number of bound states for an anion in a magnetic field. 
Some qualitative estimates can be drawn from the studying the classical 
dynamics. A general finding is that whenever the classical motion for positive energies, 
$E_{Ns}>0$, and low collective excitations $N$ have been analysed in Ref.~\cite{paper2}, the 
autodetachment process has been detected. Assuming then that the condition $E_{0s}<0$ provides the number 
of magnetically induced bound states one finds this number to depend on two parameters: the product 
of the atomic mass and the polarizability squared i.e. $\mat\kappa^2$, and the field strength $B$. In 
Figure~14 we present the domains associated with different values for the number of magnetically induced 
bound states in the ``$\mat\kappa^2 - B$'' plane. The lines separating the domains corresponding to 
one and more bound states are parallel to each other due to the same scaling behavior of the quantity 
$\E_s$ with $B$ for $s=1,2,\ldots$~: $\E_s \propto B^3$. The boundary between the domains of unbound 
and bound states is determined by the condition $E_{0s}=0$ at $s=0$ and corresponds to the different 
scaling law, $\E_0 \propto B^2$ (see Eq.~(\ref{binding_0})). It is essential to note that for every 
atom (or molecule) there exists a non-zero critical value of the field strength $\Bc$ below which the 
anion is not magnetically bound. The larger $\mat\kappa^2$ is the smaller is the critical field. 
For a given neutral system we can determine from Figure~14 the expected number of magnetically 
bound states of the anion for a given field strength. For convenience, we indicate in the figure the 
values of $\mat\kappa^2$ for some specific atoms and molecules taken from Ref.~\cite{Miller1996}. 
Particularly note worthy is the fact that some of the atoms indicated cannot bind an electron 
in the absence of a magnetic field \cite{Andersen_review}, but are shown here to form stable 
negative ions in relatively weak magnetic fields. For instance, 
$\mat\kappa^2 \approx 9\times10^5$~a.u. for the noble gas atom Ar implies that Ar$^-$ is stable 
above a critical field $\Bc \approx3 \times10^{-2}$~T. For completeness we mention that for the 
lightest atom (H atom; $\mat\kappa^2 \approx 4\times 10^4$~a.u.) to magnetically bind an electron, 
a field larger than $\Bc \approx 10$~T is needed while for the heavy Cs atom which possesses the 
largest polarisability ($\mat\kappa^2 \approx 4\times10^{10}$~a.u.) the critical field is as small 
as $\Bc \approx 10^{-6}$~T.

\subsection{Towards a quantum description of the moving magnetically induced anions}

A quantum treatment of the Hamiltonian (\ref{Heff}) turns out to be a difficult problem. 
This is in particular due to the fact that the variables $\xc$ and $\yc$, although being related to 
physically meaningful degree of freedom, correspond to a canonical pair of coordinate and 
momentum. The potential $\Veff$ given in Eq.~(\ref{Veff}) is therefore non-local in a quantum 
description of the motion of the excess electron. To proceed, basis functions have 
been introduced in Ref.~\cite{BC2003} such that the action of the effective potential on them 
can be described analytically. These functions also involve other degree of freedom related to 
the motion of the neutral species in terms of the pair $\{\XL,\YL\}$. In addition, they 
respect the integral of motion $\Lt$ given by Eq.~(\ref{L}). Two suitable sets of quantum basis 
states have been suggested in Ref.~\cite{BC2003}. We draw here attention to one of them, 
which describes the eigenstates of the Hamiltonian (\ref{Heff}) for vanishing effective potential. 
It has been called the set of {\it basis states of detached anion} and is of particular convenience for 
treating loosely bound magnetically induced anionic states. The basis states are $|u,J\rangle$ specified 
by two quantum numbers: the number $u$ takes non-negative continuous values and determines the 
kinetic energy of an isolated neutral species moving across the magnetic field, $E^{\rm at}_u = Bu/\mat$, 
while the number $J$ takes integer values which are the eigenvalues of the total longitudinal 
angular momentum $\Lt$. Expanding an eigenfunction of the Hamiltonian (\ref{Heff}) in terms of the 
basis states $|u,J\rangle$ yields an integro-differential equation, 
\be
\left[ \frac{p_z^2}{2\mu} + E^{\rm at}_u - E_J \right] \, g^{(J)}_u(z) 
= -\int_0^\infty \, {\rm d}u' \, V^{(J)}_{u',u}(z) \, g^{(J)}_{u'}(z)~.
\label{int-diff}
\ee 
Solutions of the latter equation are the ``expansion coefficients'' $g^{(J)}_u(z)$ and 
the anionic energies $E_J$ attributed to a given value $J$ of the longitudinal angular momentum $\Lt$. 
The kernel $V^{(J)}_{u',u}(z) = \langle u',J | \Veff | u,J \rangle$, which can be referred to as the 
potential functions, describes the coupling between the states of free motion of the neutral species 
across the field owing to the interaction between the species and the external electron via the 
potential $\Veff$. 

Neglecting the couplings between different states of the motion of the neutral species one can 
introduce an approximation of Eq.~(\ref{int-diff}) by a one-dimensional Schr\"odinger equation 
with a potential 
\be
W^{(J)}_u(z) = \int_0^\infty \, {\rm d}u' \, V^{(J)}_{u',u}(z)~.
\label{WJu}
\ee
Such an approximation describes the attachment of an external electron to the neutral species moving 
across a magnetic field with a specific kinetic energy $E^{\rm at}_u$. The binding energies of the 
corresponding magnetically induced states of the anion depend on the quantum numbers $u$ and $J$. 
As discussed in Ref.~\cite{BC2003}, applying this approximation can be relevant for large values of $u$. 
For $u \gg {\rm max}(1,|J|)$ it allows one to obtain the analytical estimate for the binding energies in 
form of the asymptotical expansion in a series, 
 \be
\varepsilon_{u,J} = \left(\frac{\pi}{2}\right)^2 \mu \, \frac{\kappa^2 B^3}{u^3} 
                    \left[ 1 + \frac{3}{u} \left( |J| + \frac{3}{2} \right) 
                             + \frac{6}{u^2} \left( J^2 + \frac{15}{4}\,|J|
                                                        + \frac{75}{16}\right) + \ldots \right]^2~.
\label{estimate}
\ee
The parameter $u$ in Eq.~(\ref{estimate}) can be associated with a mean kinetic energy of the neutral 
species, $\langle E^{\rm at} \rangle = Bu/\mat$. At fixed $\langle E^{\rm at} \rangle$ the states of 
moving anions are expected to possess binding energies which scale with respect to the polarizability 
and magnetic field strength according to $\kappa^2B^3$. This scaling property of the binding 
energies is essentially associated with the polarization-like behavior of the potential which 
links the external electron to the neutral species. We remind the reader that the same property 
holds for the states with $s>0$ for infinitely heavy anions, cf. Eq.~(\ref{binding_s}). 
With increase of the mean kinetic energy of neutral species, the magnetically induced moving anions 
can be naturally expected to become less bound. This is consistent with the reduction of the binding 
energies with increasing $u$ shown in Eq.~(\ref{estimate}). On the other hand, the estimates 
(\ref{estimate}) hint that the magnetically induced moving anions are expected to be more bound for 
larger absolute values of the total longitudinal angular momentum $|J|$. This finding can be particular 
important for selecting the conditions under which magnetically induced bound states or 
long-living resonances are well manifested. 

Derivation of the integro-differential equation~(\ref{int-diff}) and the detailed studies of the properties 
of potential functions $V^{(J)}_{u',u}(z)$ provided in Ref.~\cite{BC2003} open new possibilities for 
further rigorous quantum investigations of moving magnetically induced anions. Future efforts concerning 
numerical approaches to solve Eq.~(\ref{int-diff}) are expected to bring new physical results. For example, 
according to the asymptotic estimates~(\ref{estimate}) moving anions possess a sequence of magnetically 
induced bound states corresponding to different values of the total longitudinal angular momentum. 
Answering the question whether the number of these states is finite or infinite requires an accounting of the 
coupling described by the integral operator in Eq.~(\ref{int-diff}). On the other hand, the classical 
studies discussed in the previous section lead to the expectation that many magnetically induced states 
are resonance states. Eq.~(\ref{int-diff}) can be applied to study them by {\it ab initio} quantum methods. 

\section{Conclusions and Outlook}

After a classification and discussion of the binding mechanisms and properties of anions in
field-free space this review mainly focused on the anionic states which were predicted by
Avron, Herbst and Simon \cite{Avron}: In the presence of an external magnetic field of arbitrary
strength there exists an infinite manifold of bound states for any anion.
This surprising prediction stimulated a series of investigations 
\cite{BSC2000, BCS2001, paper1, paper2, BC2003} of the underlying binding mechanism and the properties of 
the unusual states of anions exposed to the magnetic field. A first analysis of the underlying
physics \cite{BSC2000} confirmed that, assuming the anions are infinitely heavy i.e. spatially fixed,
the number of the bound states induced by the magnetic field is infinite.
These states are supported by the long-range polarization-like 
attraction of the excess electron to a static neutral species along the magnetic field. 
Explicit expressions for the binding energies of the infinite series of bound states
were derived thereby obtaining two categories of scaling with respect to the field strengths:
The binding energy of the energetically lowest magnetically induced bound state scales as $\propto B^2$
whereas the series of excited states scales as $\propto B^3$.

A complete description of anions in magnetic fields has to include the motion 
of the collective (CM) degrees of freedom: In the presence of a magnetic field the CM motion
couples to the electronic motion and the corresponding interaction plays a crucial role for
the stability of the excess electron \cite{BCS2001}. 
The theoretical description of moving excited anions in magnetic fields 
requires several conceptually new methods and techniques.
These include among others canonical transformations, exploitation of the constants of 
motion as well as adiabatic separations of fast and slow degrees of freedom in the field \cite{paper1}. 
Following this route allows an identification of the prevailing effective potential which links the excess 
electron to the underlying neutral system in a moving anion. 
We have reviewed the dynamics and stability of anions governed by the latter potential
thereby elucidating the significance of the 
the coupling between the neutral species and the electron due to the anion's motion in a magnetic 
field. This coupling triggers the process of autodetachment thereby pushing some of the bound states 
for fixed nucleus into the continuum. Their number, their properties and whether they induce resonances 
in the continuum depends on the parameters of the specific anion (atomic mass and polarizability and, 
of course, also on the magnetic field strength). The latter allow for a strong variety of anionic
structures being additionally enriched by the tuning of the external parameters such as
the magnetic field strength \cite{paper2}. We have suggested a concrete approach \cite{BC2003} for the
{\it{ab initio}} quantum description of the moving anions that should be capable of addressing
all above-indicated regimes including in particular the process of motional decay. Here, much work 
is still to be done. 

Expectations in this young field of research are manifold. A rich spectrum of bound and resonance states
of the anions moving in magnetic fields is obvious by present. One should emphasize that this includes
anionic species that {\it{do not exist in the absence of the field}} (examples are Ar$^-$ and Xe$^-$).
From an experimental point of view these might be the most interesting ones to discover !
Beyond the magnetically induced polarization binding a variety of other binding mechanisms
created by the combination of magnetic and e.g. permanent dipole or quadrupole forces could
lead to other interesting anionic species with unexpected properties.

\section*{Acknowledgements}

Financial support by the Deutsche Forschungemeinschaft is gratefully acknowledged.

\clearpage

\begin{table}
\begin{center}
\caption{
 \label{tab1}
 Binding energies of the ground and first excited magnetically induced states of those atomic 
 anions which do not exist in field-free space. $\kappa$ is the polarizability of the neutral atom.}
\begin{tabular}{|l|l|l|l|l|l|}
\cline{3-6}
   \multicolumn{2}{c}{} \vline & \multicolumn{2}{c}{$B=10$~T} \vline & \multicolumn{2}{c}{$B=30$~T} \vline \\ 
\hline
  Atom  & $\kappa$,~a.u. & $\E_0$,~meV & $\E_1$,~meV & $\E_0$,~meV & $\E_1$,~meV \\
\hline
    He  & $1.38$  & $2.92\times10^{-5}$ & $4.81\times10^{-10}$ & $2.629\times10^{-4}$ & $1.29\times10^{-8}$ \\
    Be  & $37.79$ & $2.18\times10^{-2}$ & $3.59\times10^{-7}$  & $0.196$              & $9.70\times10^{-6}$ \\ 
    N   & $7.42$  & $8.41\times10^{-4}$ & $1.38\times10^{-8}$  & $7.574\times10^{-3}$ & $3.74\times10^{-7}$ \\
    Ne  & $2.67$  & $1.08\times10^{-4}$ & $1.79\times10^{-9}$  & $9.797\times10^{-4}$ & $4.84\times10^{-8}$ \\
    Mg  & $71.53$ & $7.81\times10^{-2}$ & $1.28\times10^{-6}$  & $0.703$              & $3.47\times10^{-5}$ \\
    Ar  & $11.07$ & $1.87\times10^{-3}$ & $3.08\times10^{-8}$  & $1.686\times10^{-2}$ & $8.33\times10^{-7}$ \\
    Mn  & $63.43$ & $6.14\times10^{-2}$ & $1.01\times10^{-6}$  & $0.553$              & $2.73\times10^{-5}$ \\
    Zn  & $47.9$  & $3.50\times10^{-2}$ & $5.77\times10^{-7}$  & $0.315$              & $1.55\times10^{-5}$ \\
    Si  & $16.76$ & $4.29\times10^{-3}$ & $7.07\times10^{-8}$  & $3.864\times10^{-2}$ & $1.91\times10^{-6}$ \\
    Cd  & $48.59$ & $3.60\times10^{-2}$ & $5.93\times10^{-7}$  & $0.324$              & $1.60\times10^{-5}$ \\
    Xe  & $27.29$ & $1.13\times10^{-2}$ & $1.87\times10^{-7}$  & $0.102$              & $5.06\times10^{-6}$ \\
    Hg  & $38.46$ & $2.25\times10^{-2}$ & $3.72\times10^{-7}$  & $0.203$              & $1.01\times10^{-5}$ \\
    Rn  & $35.76$ & $1.95\times10^{-2}$ & $3.21\times10^{-7}$  & $0.175$              & $8.69\times10^{-6}$ \\
\hline
\end{tabular}
\end{center}
\end{table}

\clearpage

\begin{list}{}{\leftmargin 2cm \labelwidth 1.5cm \labelsep 0.5cm}

\item[\bf Figure~1.] The total energies and detachment thresholds for the two lowest states of the 
anion H$^-$ in a magnetic field. The dots represent the results from Ref.~\cite{Strong_B_H^-} 
and are connected with lines for convenience: the solid lines represent the total energies of the 
singlet and triplet levels while the dot-dashed lines indicate the corresponding detachment thresholds. 
$1$~a.u. of the field strength corresponds to $2.3554\times10^5$~T.

\item[\bf Figure~2.] The binding energies for the singlet and triplet states of the 
anion H$^-$ in a magnetic field, according to the results from Ref.~\cite{Strong_B_H^-}. 
Both quantities are the spin-preserving binding energies, i.e., energies required to remove one 
electron from the anion without changing the longitudinal component of the total electronic spin 
(see text). 

\item[\bf Figure~3.] The true electron affinity of the hydrogen atom in a magnetic field. 
The solid curve is obtained from the results presented in Ref.~\cite{Strong_B_H^-}. Dashed 
lines show the results obtained earlier in Ref.~\cite{Surmelian1974}.

\item[\bf Figure~4.] The coordinates used to describe a moving anion as a system of a neutral atom and 
an external electron. The spatial extension corresponding to $10$~T magnetic field strength is indicated 
by the bold horizontal bar.

\item[\bf Figure~5.] Motion of the Cs$^-$ ion in terms of the four interacting degrees of freedom 
in the magnetic field $B=40$~T. 
(a) The trajectory of the atomic motion across the magnetic field for the propagation time $285$~ns. 
(b) The solid line area is the trajectory of the fast electronic motion relative to the atom in the 
plane perpendicular to the magnetic field. This trajectory is shown for the propagation time $0.5$~ns. 
The dashed line indicates the corresponding trajectory of the electronic guiding centre for a bigger 
propagation time of $7$~ns. 
(c) The electronic oscillations along the magnetic field.

\item[\bf Figure~6.] Motion of the electronic guiding centre for the trajectory shown in Figure~5: 
(a) in the plane perpendicular to the magnetic field, 
(b) the time dependence of the displacement $\rc=\sqrt{\xc^2+\yc^2}$ of the electronic guiding 
centre from the atom.

\item[\bf Figure~7.] The electronic Larmor motion corresponding to the trajectory shown in Figure~5: 
(a) in the plane perpendicular to the magnetic field (involving $1118$ cycles over the time scale 
of $0.5$~ns), 
(b) demonstrates the adiabatic conservation of the corresponding Larmor orbit radius 
$\rL=\sqrt{\xL^2+\yL^2}$.

\item[\bf Figure~8.] Stable regular motion of the Cs$^-$ ion in terms of the three 
interacting degrees of freedom (averaged over the fast electron Larmor rotations) 
in the magnetic field $B=10$~T. The energy of the anion is $-1.262\times10^{-9}$~a.u. 
given by Eq.~(\ref{Shell}) for $N=0$, $s=1$. The value of the anionic longitudinal angular 
momentum (\ref{L}) is $\Lt=0$. The top panel shows the atomic trajectory for one nearly complete 
cycle of the atomic rotation around the guiding centre corresponding to the propagation time $1000$~ns. 
The indicated displacement of the atomic guiding centre from the origin (the integral of motion) 
and the initial displacement of the atom from it correspond to those given in Figure~4. 
The bottom panel shows the rotation of the electronic guiding centre around the atom, 
and the electronic longitudinal oscillations for the propagation time $25$~ns.

\item[\bf Figure~9.] The atomic trajectories for the motion of the Cs$^-$ ion 
corresponding to the positive energy $E_{Ns}=4.820\times10^{-10}$~a.u. 
for $N=10$, $s=1$ for the magnetic field strength $B=10$~T. 
The different trajectories relate to the different values of the longitudinal angular 
momentum (\ref{L}) indicated at the arrows which point to the initial locations of the atom. 
The propagation times are about $370$~ns for the detaching trajectories and 
$1000$~ns for the non-detaching (circular-like) ones.

\item[\bf Figure~10.] The electronic motion for the selected atomic trajectories shown in Figure~9. 
The time variations of $\rc$ and $z$ are demonstrated. 
Subfigures (a), (b), (c) and (d) correspond to $\Lt=5,7,8$ and $10$, respectively.

\item[\bf Figure~11.] The fraction of the trajectories which describe the autodetachment of the extra 
electron from the anion H$^-$ due to motion-induced effects as a function of time for the magnetic 
field $B=1$~T. The initial conditions for the trajectories belong to the energy shell~(\ref{Shell}) 
at $N=s=0$, and to the value $\Lt=0$ such that at time $t=0$ the ensemble of anions corresponds 
to the energetically lowest magnetically induced bound state of the infinitely heavy anion. 
The inserts show typical trajectories corresponding to the different parts of the anionic 
autodetachment curve.

\item[\bf Figure~12.] The fractions of the detached trajectories for the Cs$^-$ anions moving in 
a magnetic field $B=1$~T with the longitudinal angular momentum $\Lt=-1$. The initial ensembles of the 
anions simulate the entire anion on the ground Landau orbit, $N=0$, but with different affinities of 
the external electron to the atom corresponding to the magnetically induced bound states 
$s=1,2,3,4$ and $5$, respectively, of the infinitely heavy ion.

\item[\bf Figure~13.] The three typical trajectories corresponding to the three different slopes of the 
autodetachment curve for $s=3$ shown in Figure~7. The top plot shows variations of the electronic 
longitudinal coordinate with time, while the bottom plot demonstrates evolutions of the transverse 
separation of the electronic guiding centre from the atom. The trajectories (a), (b) and (c) contribute 
to the correspondingly labelled parts of the autodetachment curve in Figure~12. 
Notice that the trajectories (a) and (b) for $\rc=\rc(t)$ coincide. 

\item[\bf Figure~14.] The domains of different numbers of bound magnetically induced states in the 
``$\mat\kappa^2 - B$'' plane. In this plot we also indicate the values of $\mat\kappa^2$ for some 
specific atoms and small molecules. The encircled elements are the elements which do not form a stable 
negative ion in field-free space. This figure is based on simple energetic arguments (see text). 
Quantitative corrections are expected on the ground of full quantum calculations.

\end{list}

\clearpage

\begin{figure}[ht]
  \begin{center}
   \includegraphics[width=0.95\textwidth]{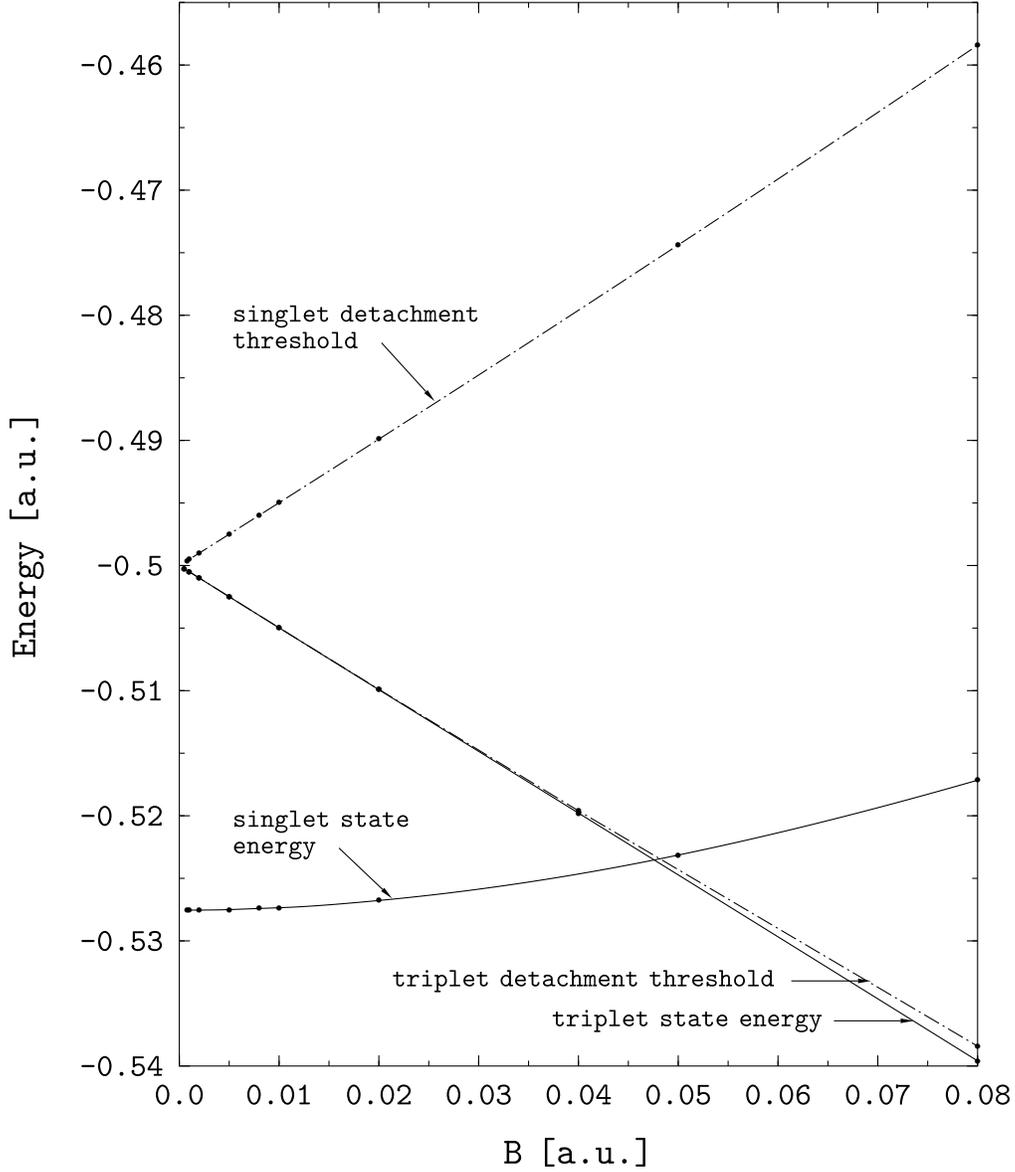}
   \caption{
The total energies and detachment thresholds for the two lowest states of the 
anion H$^-$ in a magnetic field. The dots represent the results from Ref.~\cite{Strong_B_H^-} 
and are connected with lines for convenience: the solid lines represent the total energies of the 
singlet and triplet levels while the dot-dashed lines indicate the corresponding detachment thresholds. 
$1$~a.u. of the field strength corresponds to $2.3554\times10^5$~T.
           }
  \end{center}
\end{figure}

\clearpage

\begin{figure}[ht]
  \begin{center}
   \includegraphics[width=0.95\textwidth]{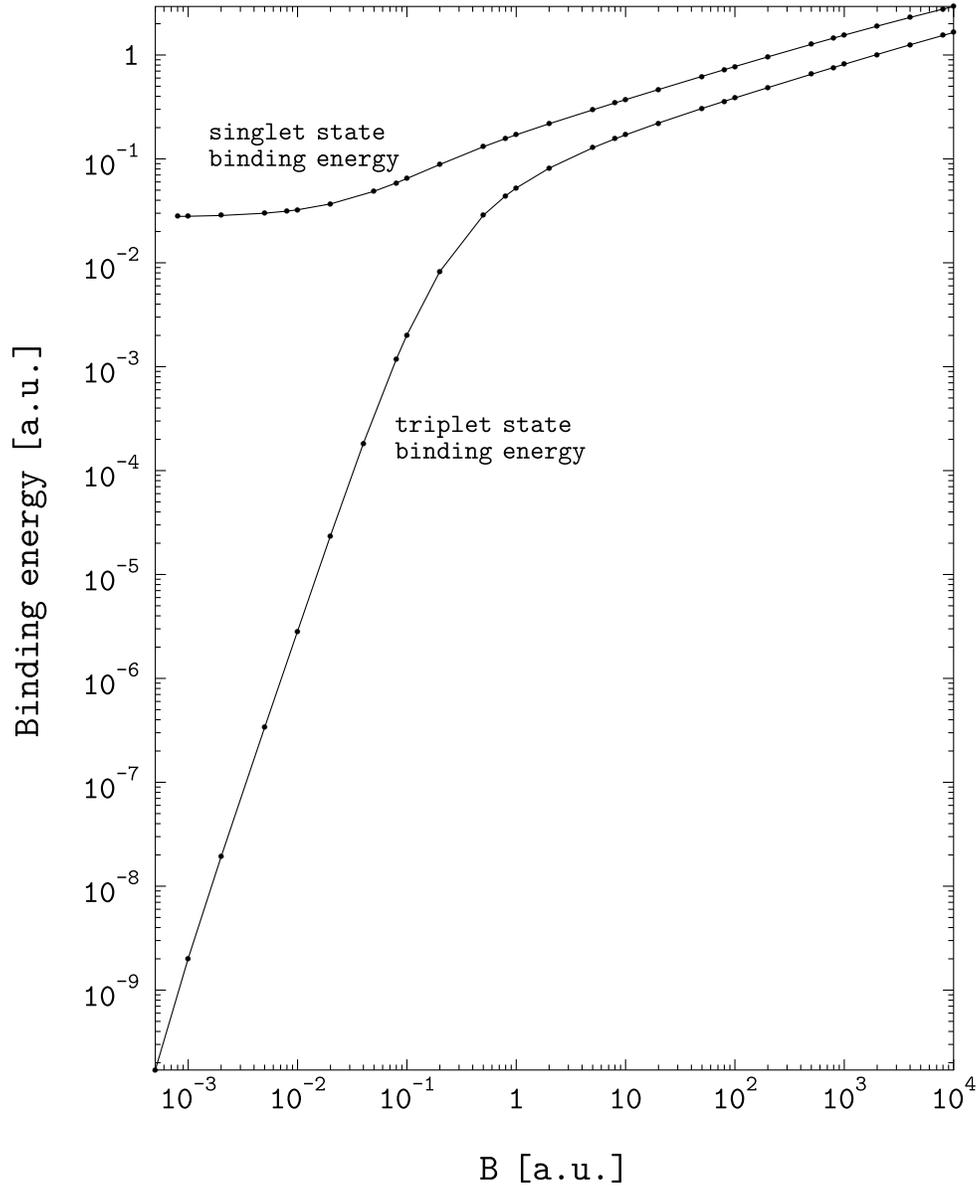}
   \caption{
 The binding energies for the singlet and triplet states of the 
anion H$^-$ in a magnetic field, according to the results from Ref.~\cite{Strong_B_H^-}. 
Both quantities are the spin-preserving binding energies, i.e., energies required to remove one 
electron from the anion without changing the longitudinal component of the total electronic spin 
(see text). 
           }
  \end{center}
\end{figure}

\clearpage

\begin{figure}[ht]
  \begin{center}
   \includegraphics[width=0.95\textwidth]{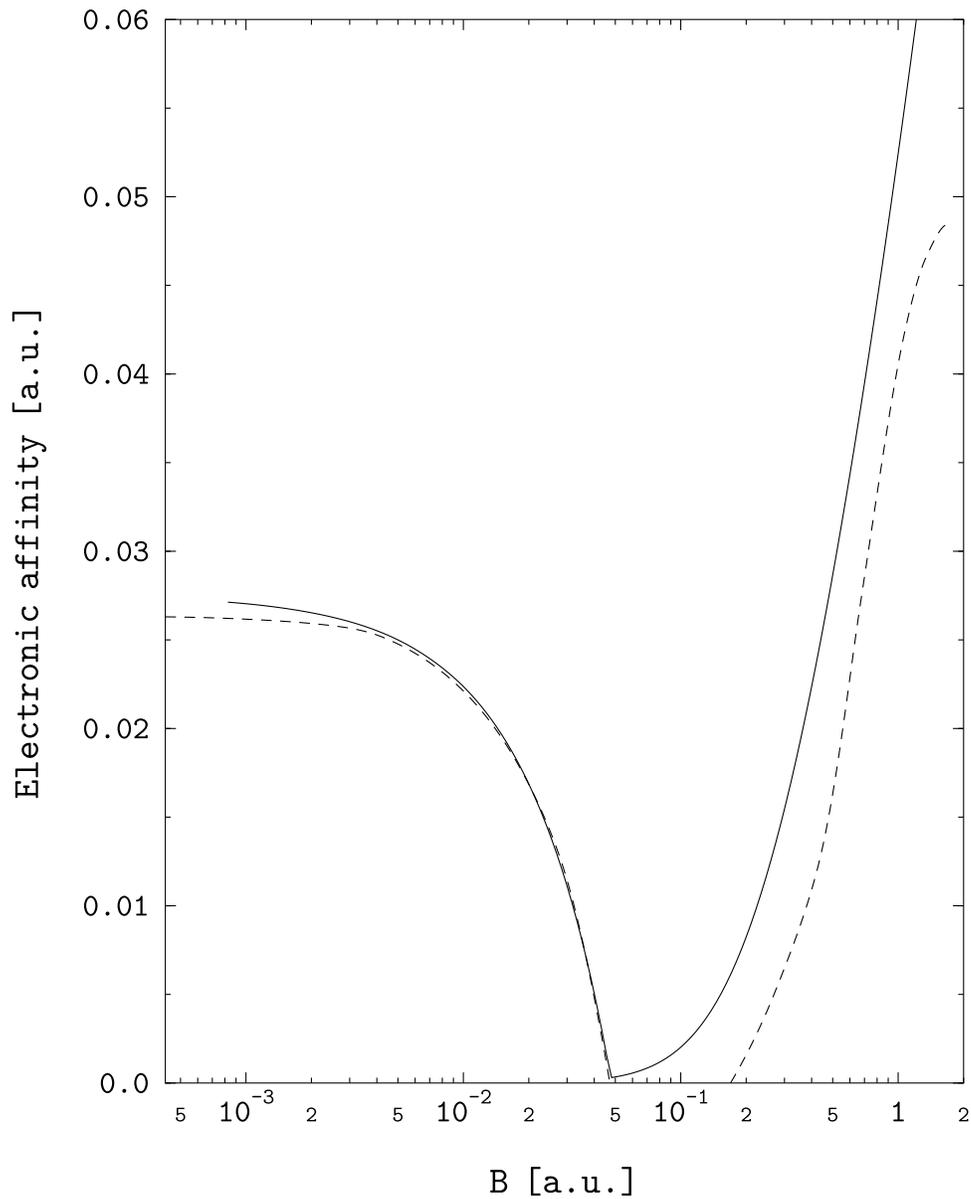}
   \caption{
The true electron affinity of the hydrogen atom in a magnetic field. 
The solid curve is obtained from the results presented in Ref.~\cite{Strong_B_H^-}. Dashed 
lines show the results obtained earlier in Ref.~\cite{Surmelian1974}.
           }
  \end{center}
\end{figure}

\clearpage

\begin{figure}[ht]
  \begin{center}
   \includegraphics[width=0.95\textwidth]{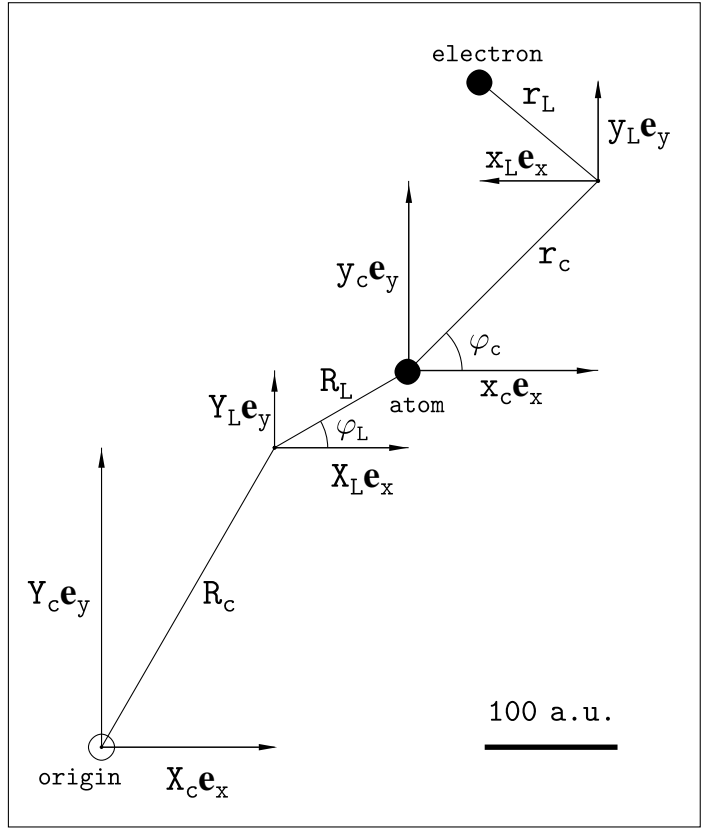}
   \caption{
The coordinates used to describe a moving anion as a system of a neutral atom and 
an external electron. The spatial extension corresponding to $10$~T magnetic field strength is indicated 
by the bold horizontal bar.
           }
  \end{center}
\end{figure}

\clearpage

\begin{figure}[ht]
  \begin{center}
   \includegraphics[width=0.50\textwidth]{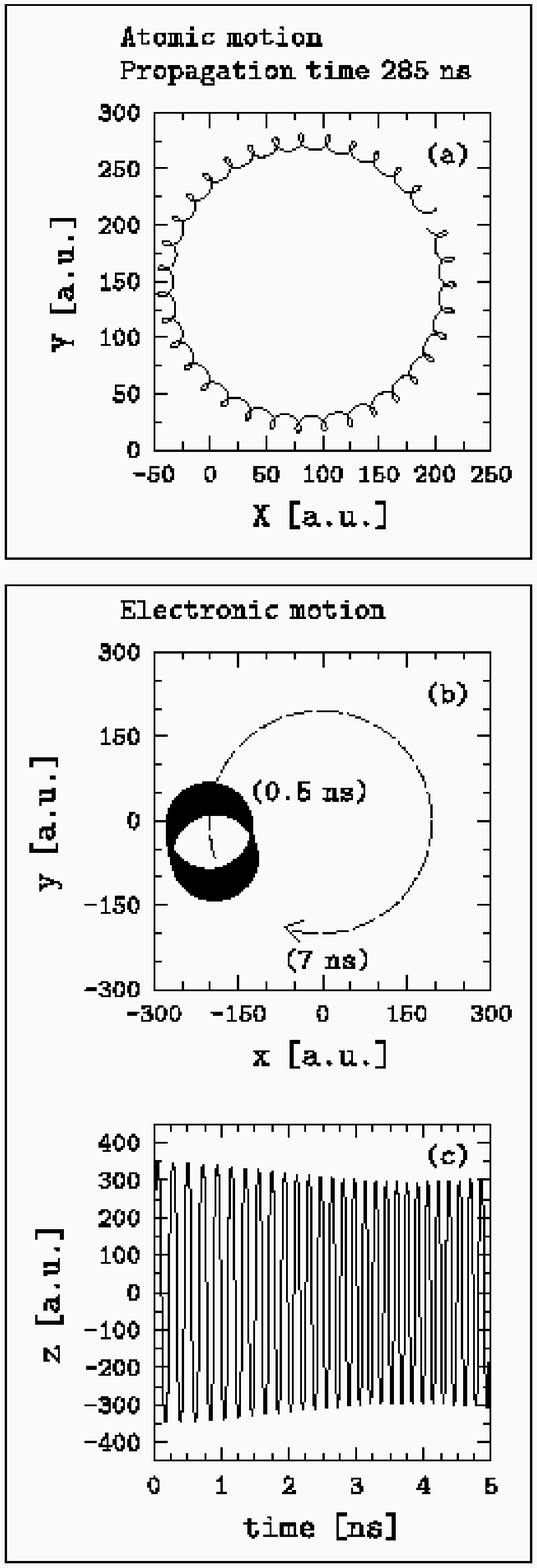}
   \caption{
Motion of the Cs$^-$ ion in terms of the four interacting degrees of freedom 
in the magnetic field $B=40$~T. 
(a) The trajectory of the atomic motion across the magnetic field for the propagation time $285$~ns. 
(b) The solid line area is the trajectory of the fast electronic motion relative to the atom in the 
plane perpendicular to the magnetic field. This trajectory is shown for the propagation time $0.5$~ns. 
The dashed line indicates the corresponding trajectory of the electronic guiding center for a bigger 
propagation time of $7$~ns. 
(c) The electronic oscillations along the magnetic field.
           }
  \end{center}
\end{figure}

\clearpage

\begin{figure}[ht]
  \begin{center}
   \includegraphics[width=0.85\textwidth]{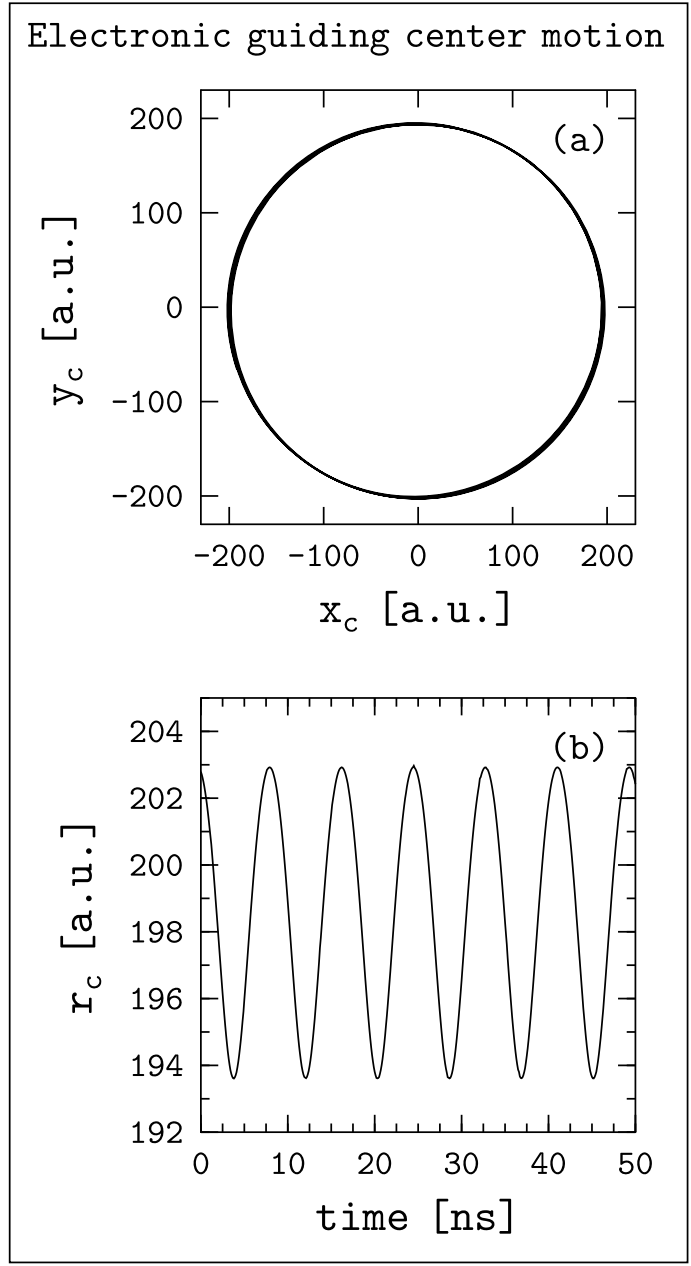}
   \caption{
Motion of the electronic guiding centre for the trajectory shown in Figure~5: 
(a) in the plane perpendicular to the magnetic field, 
(b) the time dependence of the displacement $\rc=\sqrt{\xc^2+\yc^2}$ of the electronic guiding 
centre from the atom.
           }
  \end{center}
\end{figure}

\clearpage

\begin{figure}[ht]
  \begin{center}
   \includegraphics[width=0.85\textwidth]{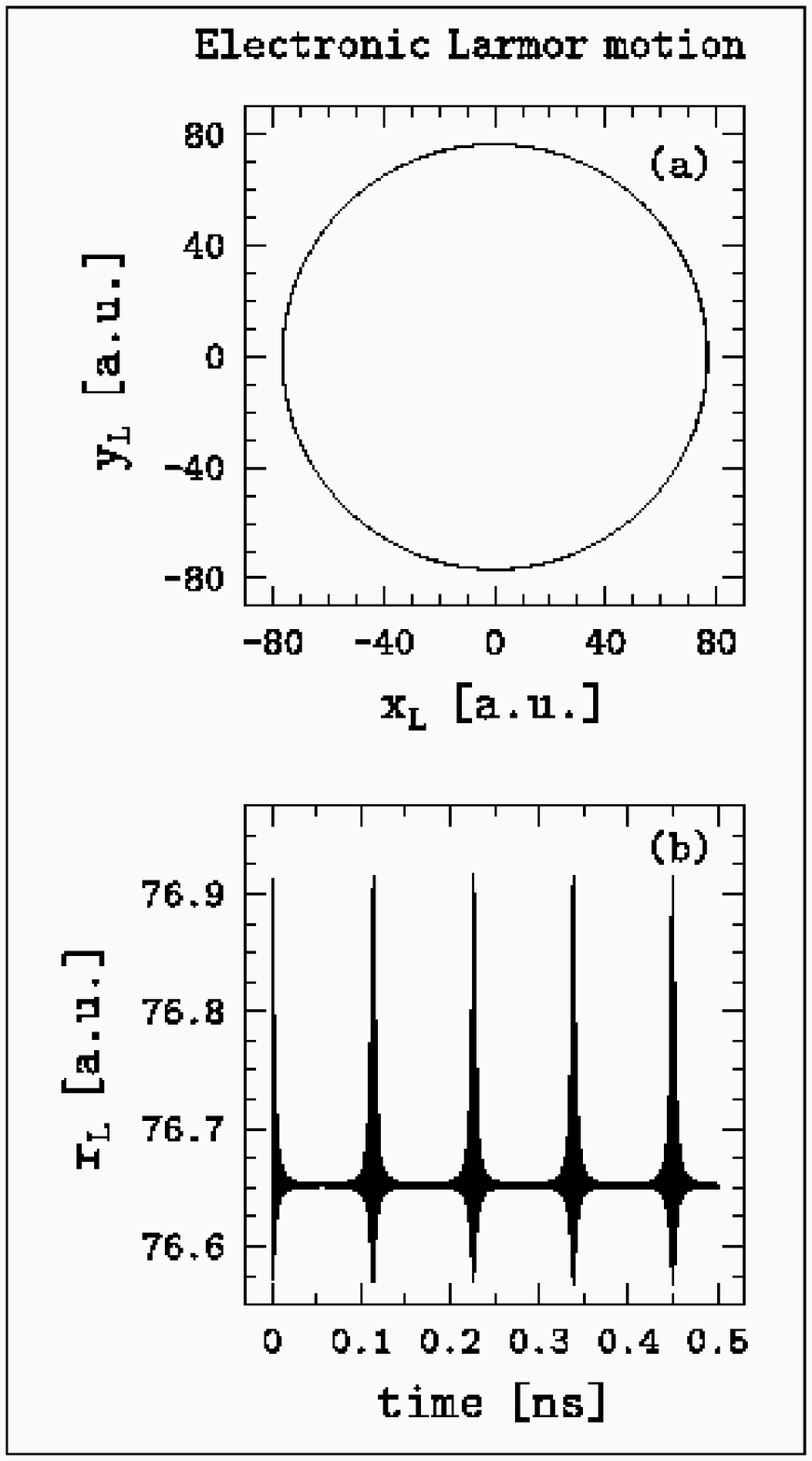}
   \caption{
The electronic Larmor motion corresponding to the trajectory shown in Figure~5: 
(a) in the plane perpendicular to the magnetic field (involving $1118$ cycles over the time scale 
of $0.5$~ns), 
(b) demonstrates the adiabatic conservation of the corresponding Larmor orbit radius 
$\rL=\sqrt{\xL^2+\yL^2}$.
           }
  \end{center}
\end{figure}

\clearpage

\begin{figure}[ht]
  \begin{center}
   \includegraphics[width=0.43\textwidth]{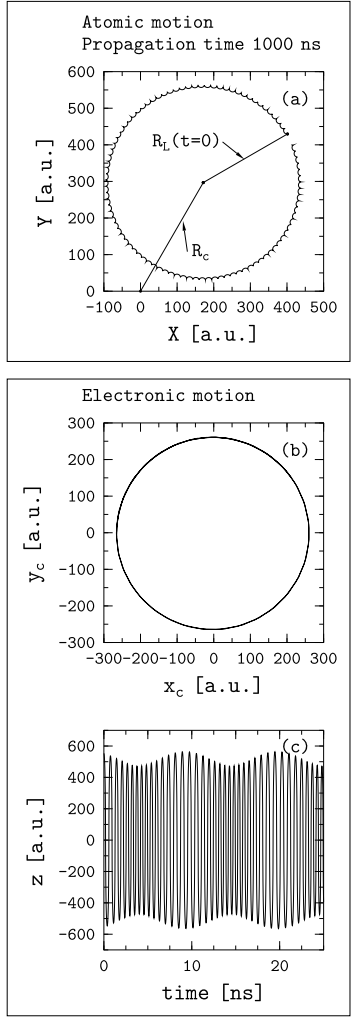}
   \caption{
Stable regular motion of the Cs$^-$ ion in terms of the three 
interacting degrees of freedom (averaged over the fast electron Larmor rotations) 
in the magnetic field $B=10$~T. The energy of the anion is $-1.262\times10^{-9}$~a.u. 
given by Eq.~(\ref{Shell}) for $N=0$, $s=1$. The value of the anionic longitudinal angular 
momentum (\ref{L}) is $\Lt=0$. The top panel shows the atomic trajectory for one nearly complete 
cycle of the atomic rotation around the guiding centre corresponding to the propagation time $1000$~ns. 
The indicated displacement of the atomic guiding centre from the origin (the integral of motion) 
and the initial displacement of the atom from it correspond to those given in Figure~4. 
The bottom panel shows the rotation of the electronic guiding centre around the atom, 
and the electronic longitudinal oscillations for the propagation time $25$~ns.
           }
  \end{center}
\end{figure}

\clearpage

\begin{figure}[ht]
  \begin{center}
   \includegraphics[width=0.95\textwidth]{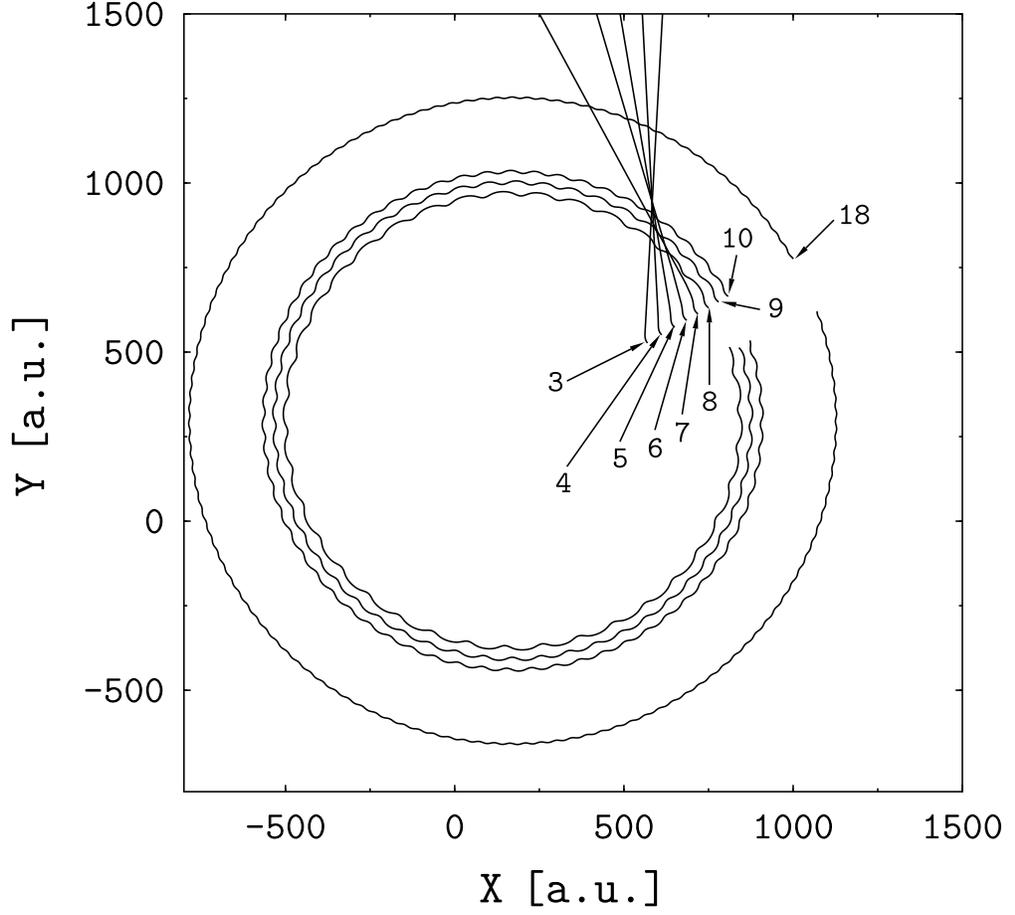}
   \caption{
The atomic trajectories for the motion of the Cs$^-$ ion 
corresponding to the positive energy $E_{Ns}=4.820\times10^{-10}$~a.u. 
for $N=10$, $s=1$ for the magnetic field strength $B=10$~T. 
The different trajectories relate to the different values of the longitudinal angular 
momentum (\ref{L}) indicated at the arrows which point to the initial locations of the atom. 
The propagation times are about $370$~ns for the detaching trajectories and 
$1000$~ns for the non-detaching (circular-like) ones.
           }
  \end{center}
\end{figure}

\clearpage

\begin{figure}[ht]
  \begin{center}
   \includegraphics[width=0.58\textwidth]{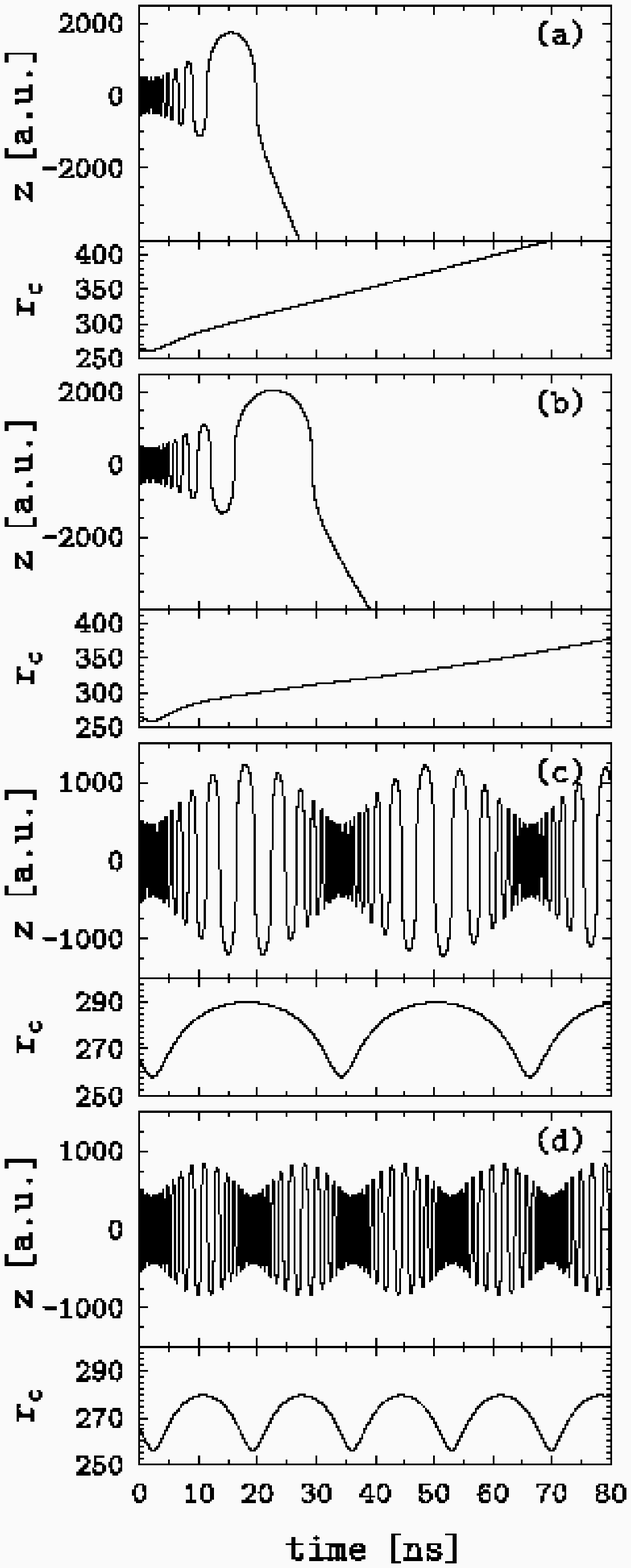}
   \caption{
The electronic motion for the selected atomic trajectories shown in Figure~9. 
The time variations of $\rc$ and $z$ are demonstrated. 
Subfigures (a), (b), (c) and (d) correspond to $\Lt=5,7,8$ and $10$, respectively.
           }
  \end{center}
\end{figure}

\clearpage

\begin{figure}[ht]
  \begin{center}
   \includegraphics[width=0.90\textwidth]{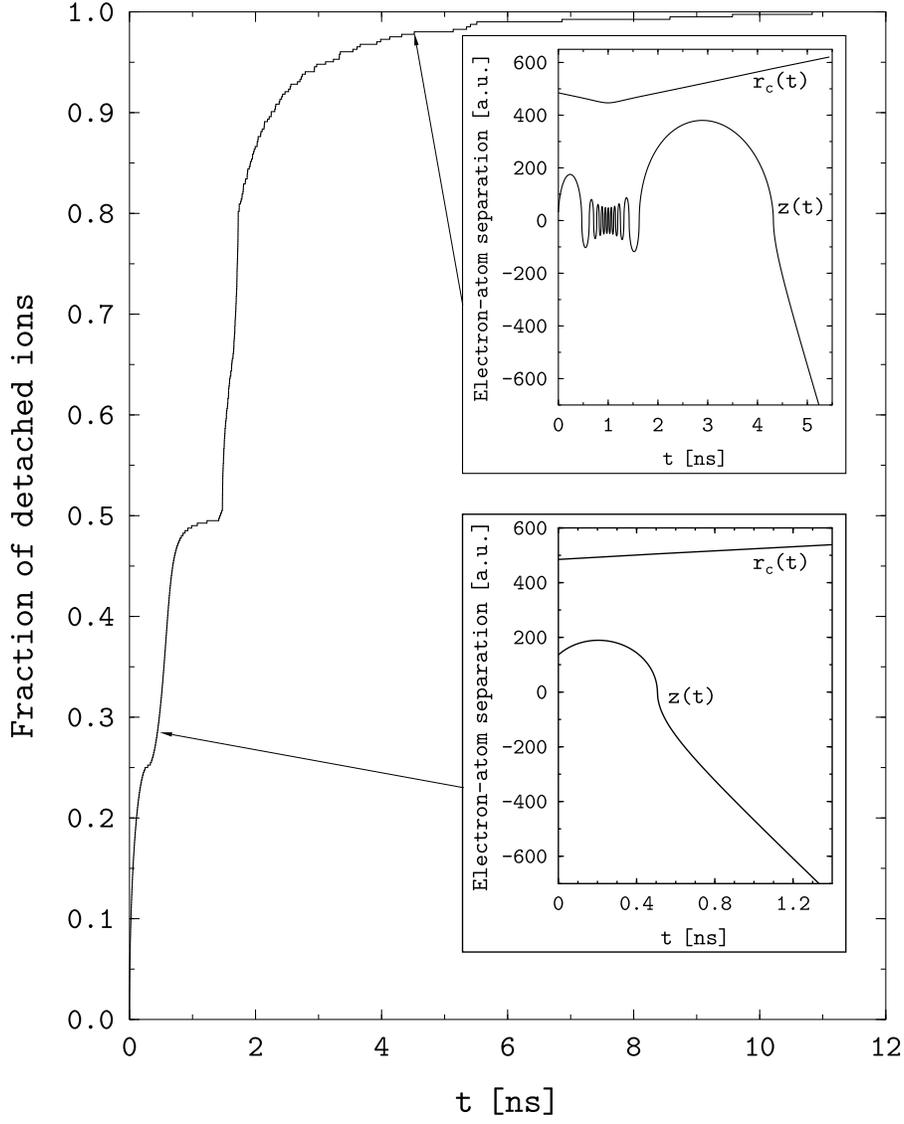}
   \caption{
The fraction of the trajectories which describe the autodetachment of the extra 
electron from the anion H$^-$ due to motion-induced effects as a function of time for the magnetic 
field $B=1$~T. The initial conditions for the trajectories belong to the energy shell~(\ref{Shell}) 
at $N=s=0$, and to the value $\Lt=0$ such that at time $t=0$ the ensemble of anions corresponds 
to the energetically lowest magnetically induced bound state of the infinitely heavy anion. 
The inserts show typical trajectories corresponding to the different parts of the anionic 
autodetachment curve.
           }
  \end{center}
\end{figure}

\clearpage

\begin{figure}[ht]
  \begin{center}
   \includegraphics[width=0.95\textwidth]{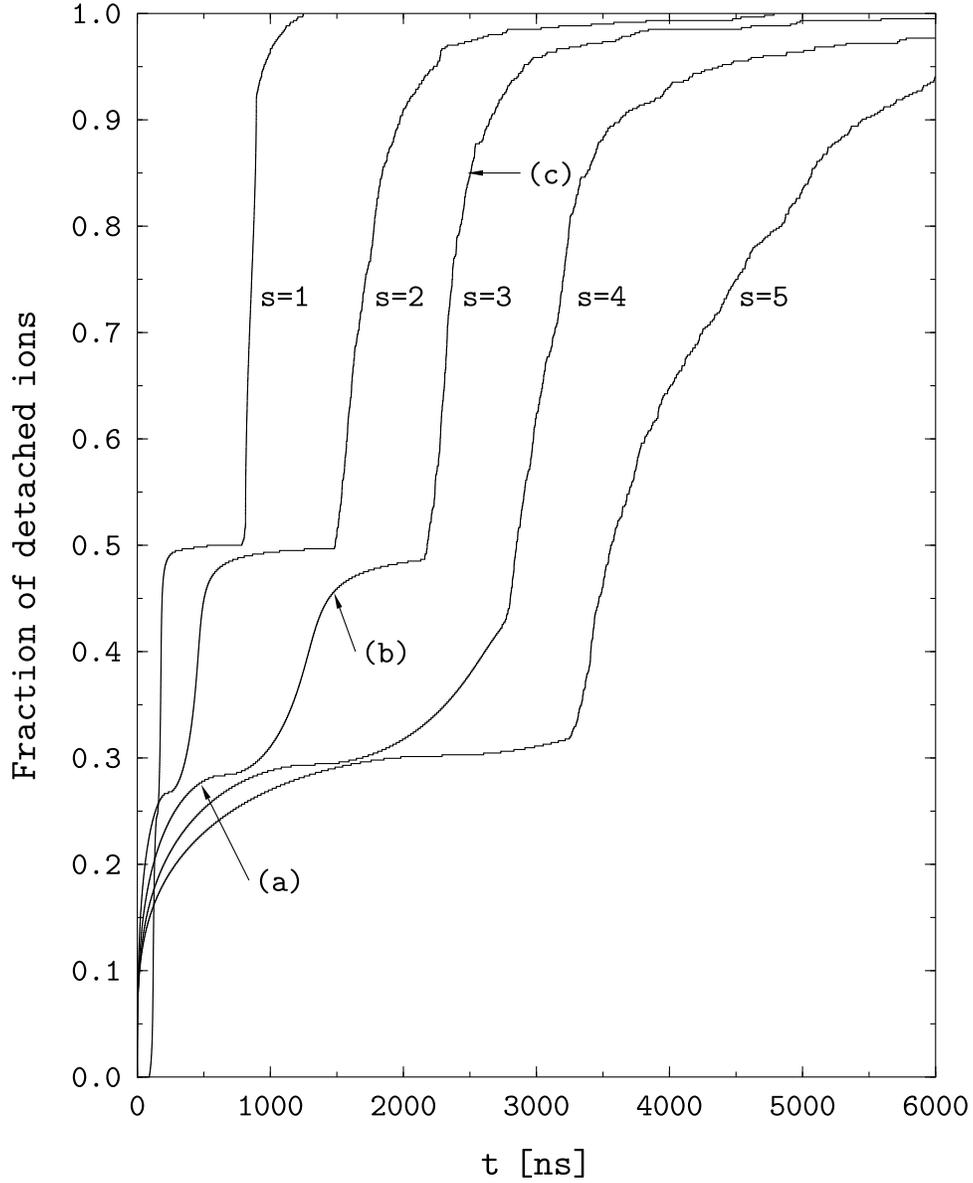}
   \caption{
The fractions of the detached trajectories for the Cs$^-$ anions moving in 
a magnetic field $B=1$~T with the longitudinal angular momentum $\Lt=-1$. The initial ensembles of the 
anions simulate the entire anion on the ground Landau orbit, $N=0$, but with different affinities of 
the external electron to the atom corresponding to the magnetically induced bound states 
$s=1,2,3,4$ and $5$, respectively, of the infinitely heavy ion.
           }
  \end{center}
\end{figure}

\clearpage

\begin{figure}[ht]
  \begin{center}
   \includegraphics[width=0.78\textwidth]{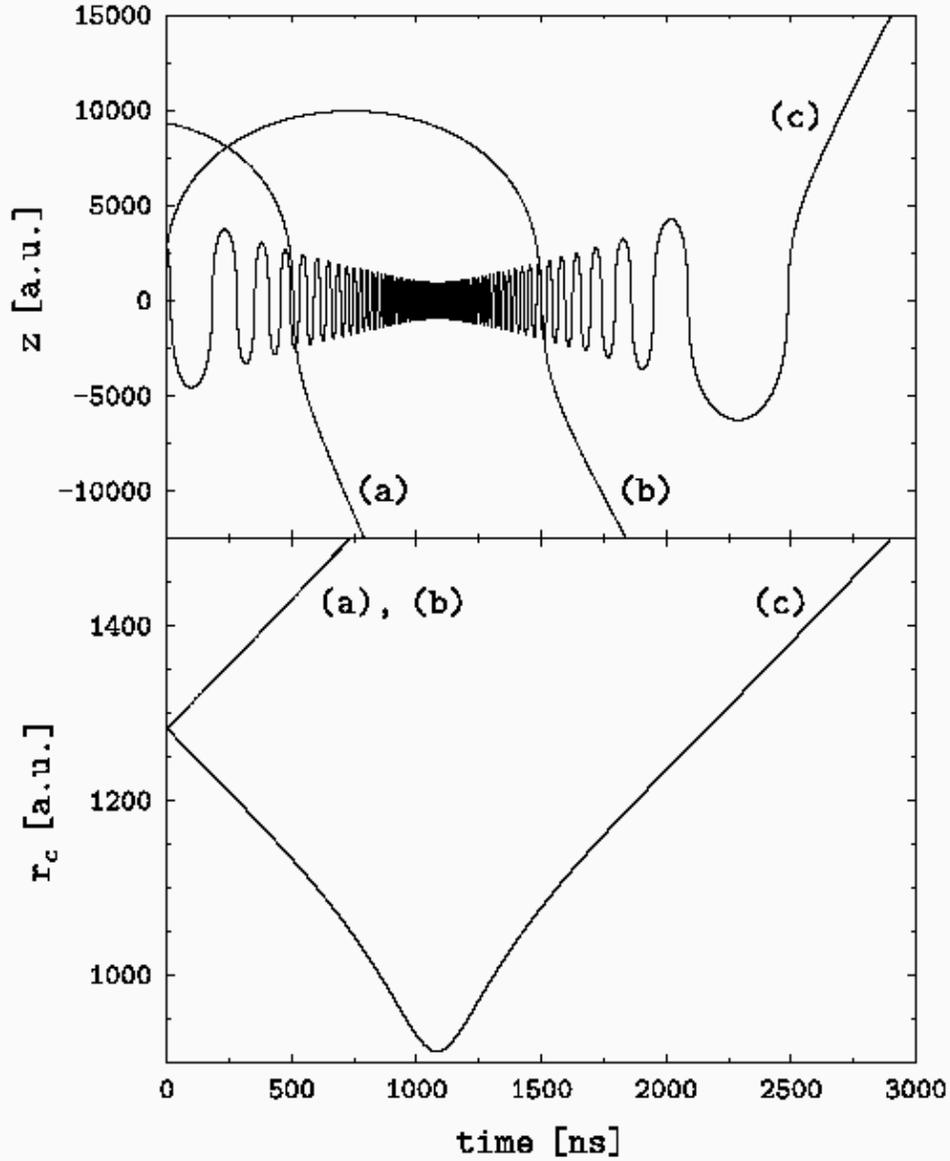}
   \caption{
The three typical trajectories corresponding to the three different slopes of the 
autodetachment curve for $s=3$ shown in Figure~7. The top plot shows variations of the electronic 
longitudinal coordinate with time, while the bottom plot demonstrates evolutions of the transverse 
separation of the electronic guiding centre from the atom. The trajectories (a), (b) and (c) contribute 
to the correspondingly labelled parts of the autodetachment curve in Figure~12. 
Notice that the trajectories (a) and (b) for $\rc=\rc(t)$ coincide. 
           }
  \end{center}
\end{figure}

\clearpage

\begin{figure}[ht]
  \begin{center}
   \includegraphics[width=0.80\textwidth]{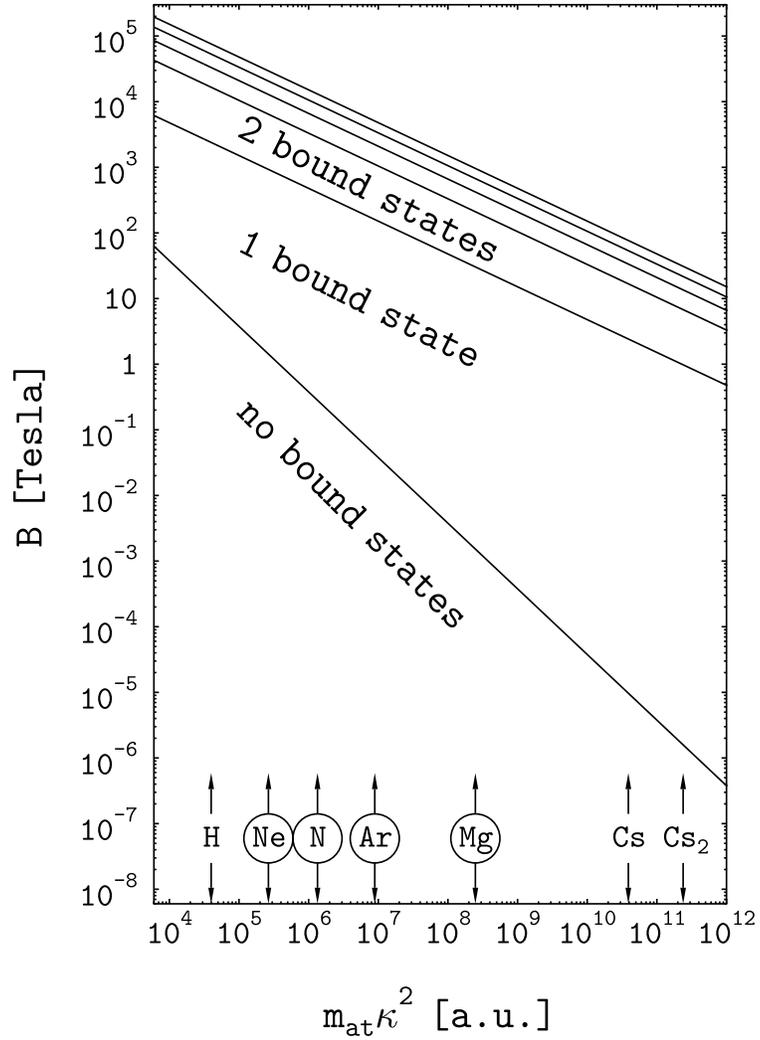}
   \caption{
The domains of different numbers of bound magnetically induced states in the 
``$\mat\kappa^2 - B$'' plane. In this plot we also indicate the values of $\mat\kappa^2$ for some 
specific atoms and small molecules. The encircled elements are the elements which do not form a stable 
negative ion in field-free space. This figure is based on simple energetic arguments (see text). 
Quantitative corrections are expected on the ground of full quantum calculations.
           }
  \end{center}
\end{figure}

\end{document}